\documentclass[manuscript,screen,acmsmall]{acmart}

\usepackage[T1]{fontenc}
\usepackage{tipa}
\usepackage{booktabs, multirow}
\usepackage{makecell}
\usepackage{xcolor,colortbl} 
\usepackage{adjustbox}
\usepackage{changepage,threeparttable} 
\usepackage{tcolorbox}
\usepackage{amsmath}
\usepackage{tabularx}
\usepackage{enumitem}
\usepackage{url}
\usepackage{verbatim}
\usepackage{graphicx}
\usepackage{subcaption}
\usepackage{tikz}
\usepackage{styles/pgf-pie}
\usetikzlibrary{positioning,shadows}
\newif\ifpienumberinlegend
\pgfkeys{/number in legend/.code=
    \expandafter\let\expandafter\ifpienumberinlegend
    \csname if#1\endcsname
    \ifpienumberinlegend

    \def\beforenumber##1\afternumber{}%
    \fi,
    /number in legend/.default=true
}

\usepackage{pifont}
\newcommand{\cmark}{\ding{51}}%
\newcommand{\xmark}{\ding{55}}%

\usepackage{tcolorbox}
\newcommand{\rqboxc}[1]{\begin{tcolorbox}[left=1pt,right=1pt,top=0pt,bottom=0pt,colback=gray!5,colframe=gray!40!black,before skip=5pt,after skip=5pt] #1\end{tcolorbox}}

\AtBeginDocument{%
  }

\begin{document}

\title{Optimized Log Parsing with Syntactic Modifications}

\author{Nafid Enan}
\email{enan@yorku.ca}
\orcid{0009-0007-4357-5719}
\author{Gias Uddin}
\email{guddin@yorku.ca}
\orcid{0000-0003-1376-095X}
\affiliation{%
  \institution{York University}
  \city{Toronto}
  \state{Ontario}
  \country{Canada}
}


\begin{abstract}

  Logs provide valuable insights into system runtime and assist in software
  development and maintenance.  Log parsing, which converts semi-structured log
  data into structured log data, is often the first step in automated log
  analysis. Given the wide range of log parsers utilizing diverse techniques, it
  is essential to evaluate them to understand their characteristics and
  performance. In this paper, we conduct a comprehensive empirical study
  comparing syntax- and semantic-based log parsers, as well as single-phase and
  two-phase parsing architectures. Our experiments reveal that semantic-based
  methods perform better at identifying the correct templates and syntax-based
  log parsers are 10 to 1,000 times more efficient and provide better grouping
  accuracy although they fall short in accurate template identification.
  Moreover, two-phase architecture consistently improves accuracy compared to
  single-phase architecture. Based on the findings of this study, we propose
  SynLog+, a template identification module that acts as the second phase in a
  two-phase log parsing architecture. SynLog+ improves the parsing accuracy of
  syntax-based and semantic-based log parsers by 236\% and 20\% on average,
  respectively, with virtually no additional runtime cost.
  Replication package.
  \url{https://github.com/disa-lab/SynLogPlus}

\end{abstract}




\maketitle



\section{Introduction}

Software logs provide critical insights into runtime behavior, errors, and
failures, but their volume and complexity make manual analysis impractical. To
assist in log analysis, researchers have proposed a number of approaches to
automate tasks such as anomaly detection~\cite{DeepLog, LogDAPT, LogFit},
failure prediction~\cite{FailPredictFronza, FailPredictDas}, and root cause
analysis~\cite{SherLog, RCALu}.  In automating log analysis, log parsing often
serves as the first step.

\begin{figure}[h]
  \centering
  \includegraphics[width=1\columnwidth]{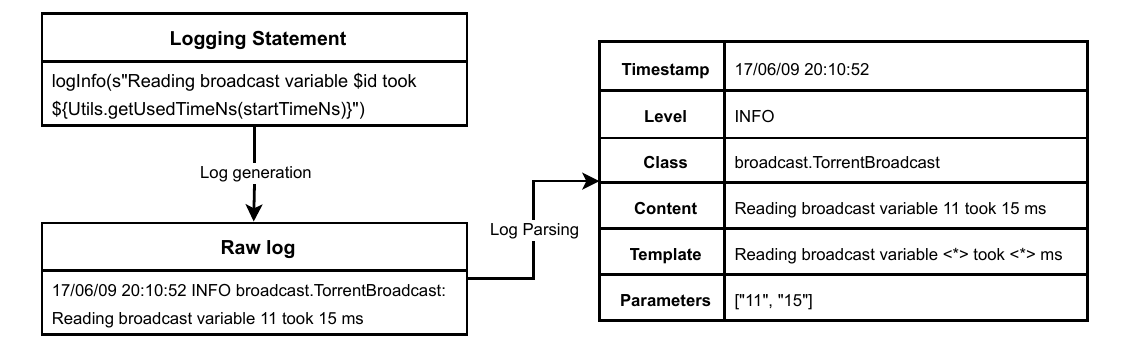}
  \caption{An example of log parsing from Spark dataset}
  \label{fig: parsing}
\end{figure}

The aim of log parsing is to convert the raw log data into log templates by
identifying the constants and variables.  The logging statement in the source
code contains both constants and dynamic variables.  Log parsing converts the
log content into a log template which delineates between the constants and the
variables in the logging statement. For instance, in Figure~\ref{fig: parsing},
the log content ``\textit{Reading broadcast variable 11 took 15 ms}'' is parsed
into the log template ``\textit{Reading broadcast variable <*> took <*> ms}''
with template parameters \textit{11} and \textit{15} where the parameters are
the values of the variables in the log template.

Since it is a fundamental step for downstream log analysis tasks, extensive
research has been done in log parsing.  Initially, the research focused on
syntax-based log parsers, which utilized frequency, similarity, and heuristics
to cluster similar log messages and identify the templates of the log
groups~\cite{LFA, SHISO, Drain}.  However, with the advances in deep learning
and language model techniques in recent years, the research in log parsing has
shifted to utilize semantic-based approaches~\cite{LogPPT, LogPrompt}, which
formulate the task of log parsing as a token classification problem.

Besides the techniques used to assess the underlying token structures, log
parsers can also be categorized based on their architectural design. Traditional
approaches often adopt a single-phase architecture, treating log parsing as a
unified task of either grouping similar messages or classifying tokens as
constants or variables. However, due to the high cost of runtime for token
classification models, recent log parsers adopt a two-phase architecture,
integrating log grouping and token classification tasks with two separate
modules~\cite{LILAC}. This reduces the inference cost of semantic models and
improves grouping accuracy by incorporating global information.

Due to the diversity of existing log parsers, it is necessary to evaluate and
analyze them not only by their parsing techniques but also by their
architectural design. Zhu et al.~\cite{Loghub} conducted the first comparative
study of 13 log parsers using a benchmark of 16 datasets, each containing 2,000
log messages (Loghub-2k). More recently, Jiang et al.~\cite{Loghub2} proposed an
evaluation study of 15 log parsers on a benchmark of 14 datasets each containing
an average of 200K logs.  However, the aforementioned studies neglected
LLM-based log parsers such as LLMParser and LILAC~\cite{LLMParser, LILAC}.
Moreover, they only evaluated the log parsing techniques.  It is equally
important to evaluate the log parsers based on their underlying architectures.

In this paper, we present a comprehensive empirical study comparing log parsing
techniques of state-of-the-art log parsers, focusing on both syntax-based and
semantic-based approaches, along with single- and two-phase log parsing
architectures. Our results show that syntax-based log parsing techniques achieve
superior grouping accuracy and runtime efficiency, while semantic-based log
parsing techniques lead in parsing accuracy but incur significantly higher
computational cost. Specifically, AEL~\cite{AEL} and Drain~\cite{Drain} achieve
a grouping accuracy (GA) of 0.8 and a parsing accuracy (PA) of 0.4, while
LogPPT~\cite{LogPPT} and LLMParser~\cite{LLMParser} achieve GA of 0.7 and PA of
0.8. LILAC achieves the best grouping and parsing accuracy while retaining
better efficiency than other semantic-based log parsers by utilizing two-phase
parsing architecture.  The experiments demonstrate that compared to single-phase
parsing, two-phase parsing obtains better accuracy across all four metrics, with
average improvement of 27\% in GA, 6\% in PA, 49\% in FGA, and 81\% in FTA.
These findings underscore the practical trade-offs between efficiency and
accuracy in syntax- and semantic-based log parsing techniques and highlight the
effectiveness of two-phase architectures.

Based on the findings of the empirical study, we introduce SynLog+, a two-phase
template identification pipeline.  SynLog+ improves the parsing accuracy of
syntax-based log parsers by 236\% on average while retaining their efficiency.
The improvement in PA for semantic-based parsers is low, an average of 20.6\%,
which is because the parsing accuracies of semantic-based log parsers are
already close to their grouping accuracies.  The average improvement in GA, PA,
FGA, and FTA achieved by employing SynLog+ is 17\%, 157\%, 181\%, and 553\%.


\section{Background}

\begin{figure}[t]
  \centering
  \includegraphics[width=0.8\columnwidth]{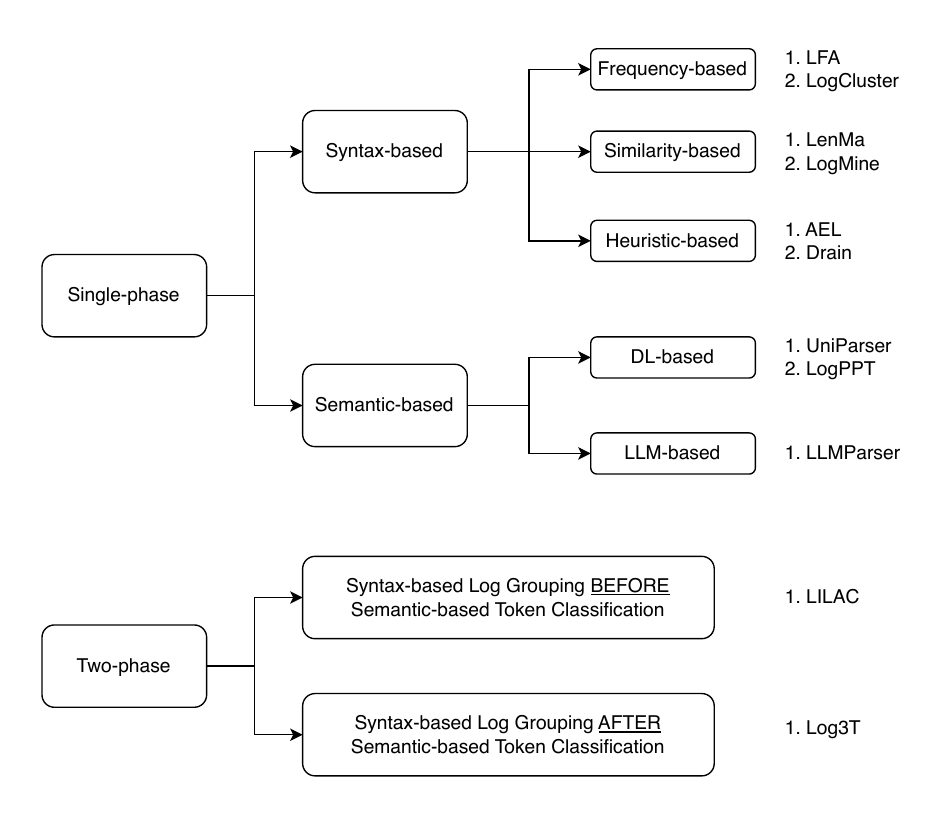}
  \caption{Hierarchical categorization of log parsers}
  \label{fig: app}
  \vspace{-6pt}
\end{figure}


\begin{figure}[t]
  \centering
  \includegraphics[width=0.9\columnwidth]{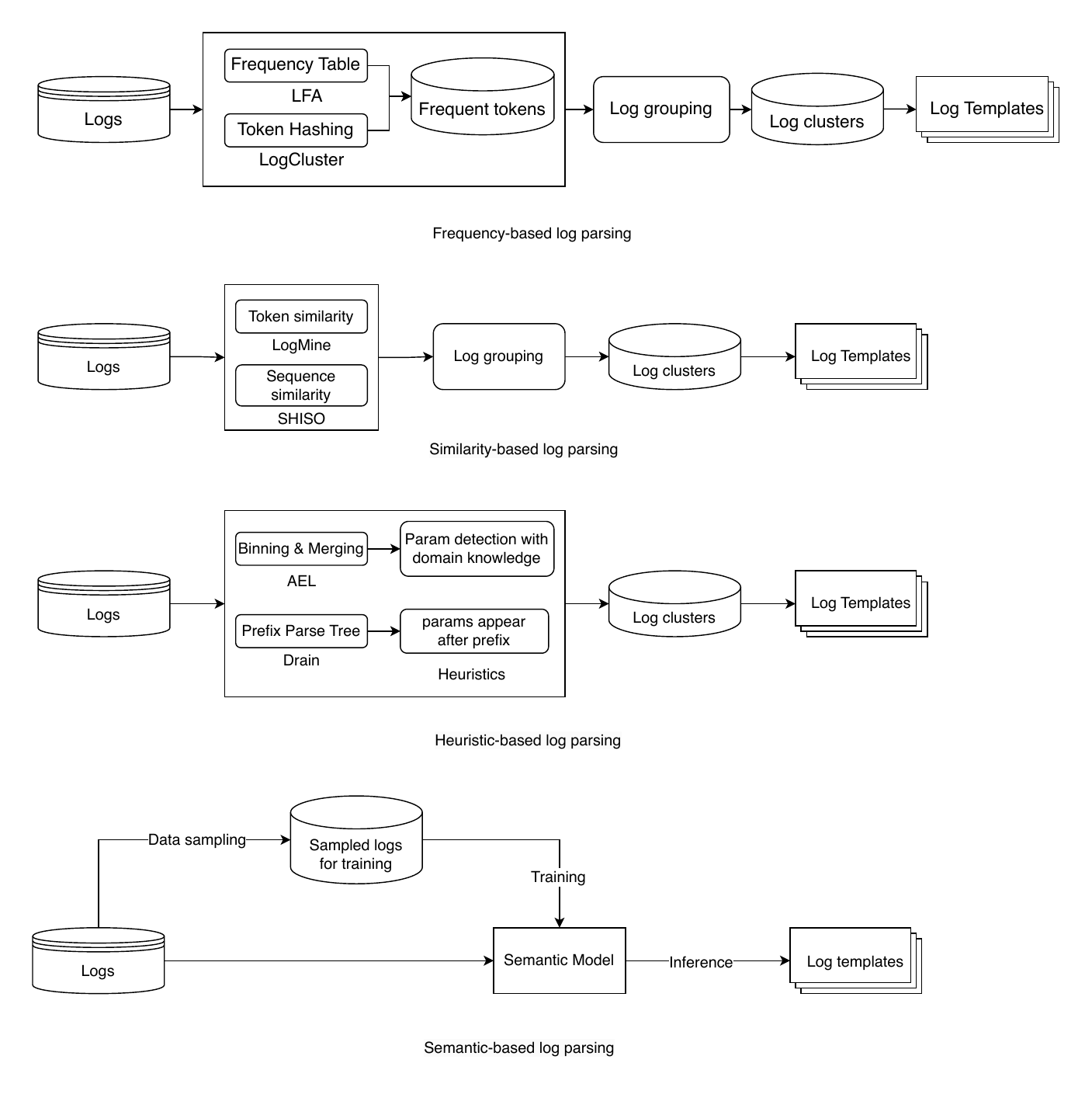}
  \caption{Overview of syntax-based and semantic-based log parsers}
  \label{fig: syn}
\end{figure}


Log parsers can be categorized in two ways: 1) based on architectures, and 2)
based on analysis of tokens.  Figure~\ref{fig: app} presents a hierarchical
categorization of the existing log parsers.

\subsection{Log parsing techniques}

Based on how the log parsers assess the underlying structure of the tokens in a
log message, existing log parsing techniques can be categorized into two
categories: (1) syntax-based and (2) semantic-based.  Figure~\ref{fig: syn}
shows the overview of syntax- and semantic-based log parsing techniques.

\subsubsection{Syntax-based log parsers}

Formulating the log parsing task as a clustering problem, Syntax-based log
parsers rely on statistical features like token frequency, position, and
patterns to group similar logs together and identify the template of the log groups.
Syntax-based methods can be divided into three types: (a) frequency-based,
(b) similarity-based, and (c) heuristic-based.

\textit{Frequency-based log parsers} operate under the assumption that constant
tokens in log messages occur more frequently than variable tokens. Based on this
assumption, they employ pattern mining techniques to identify frequent tokens.
Three prominent examples of this category are LFA, LogCluster, and
Logram~\cite{LFA, LogCluster, Logram}.

\textit{Similarity-based log parsers} assume that log messages generated from
the same event template share similar sequences of tokens.  Logs are clustered
based on the similarity between each pair of logs. LenMa~\cite{LenMA},
LogMine~\cite{LogMine}, and SHISO~\cite{SHISO} are three examples of this
category.

\textit{Heuristic-based log parsers} utilize rule-driven methods that rely on
assumptions about the log messages.  For example, Drain relies on the heuristic
assumption that the first tokens of logs are constants. Three examples of this
category are AEL, Spell, and Drain~\cite{AEL, Spell, Drain}.

\subsubsection{Semantic-based log parsers}

Instead of the syntactical structure of the tokens, semantic-based log parsers
leverage deep learning models or pretrained language models to learn the
contextual meaning of tokens. They identify the log template by using the model
to classify the tokens as either constants or variables.
We further categorize semantic-based methods into
two types: (a) Deep Learning (DL)-based, and (b) Large Language Model
(LLM)-based.

\textit{DL-based log parsers} aim to automatically learn the structural patterns
and semantic relationships within log messages using neural models. Two examples
of DL-based log parsers are UniParser~\cite{UniParser} and LogPPT~\cite{LogPPT}.

\textit{LLM-based log parsers} use LLM for token classification, either training
it with fine-tuning or prompting it with in-context learning. An example of
LLM-based log parser is LLMParser~\cite{LLMParser}.

\subsubsection{Key differences}

While syntax-based parsers employ statistical and heuristic techniques,
semantic- based parsers use various ML or LLMs to learn the semantic features of
the log data.  Moreover, syntax-based log parsers formulate the task as a
clustering problem whereas semantic-based log parsers formulate the task as a
token classification problem.

\subsection{Log parsing architectures}

Based on the underlying architecture used to identify the log templates, log
parsers can be categorized into two categories: 1) single-phase and 2)
two-phase.

\subsubsection{Single-phase log parsing}

This architecture formulates log parsing as a single problem, either log
clustering or token classification. It performs parsing in one unified phase
without separating the grouping and template identification phases. For example,
Drain and AEL cluster logs based on global patterns, achieving high grouping
accuracy but poor template identification.

\subsubsection{Two-phase log parsing}

This architecture modularizes log grouping and template
identification phases in log parsing. For instance, Log3T uses a BERT-based
classifier to identify constant tokens and matches logs to templates based on
these tokens, combining template identification and grouping phases.

\subsubsection{Key differences}

Single-phase architectures formulate the task of log parsing as either a
clustering problem or a token classification problem.  On the other hand,
two-phase architectures modularize the task of log parsing and treat both
objectives, correct grouping and correct identification of constants and
variables, with equal significance.

\section{Empirical Study}\label{sec: emp}


To understand the strengths of current log parsing tools, we
conduct an empirical study comparing syntax-based and semantic-based log
parsers, as well as single-phase and two-phase parsing architectures.

\noindent\textbf{Category 1.} Log Parser Performance.
\begin{enumerate}[label=\textbf{RQ\arabic{*}.}, start=1]

    \item How do syntax-based log parsers perform compared to semantic-based log parsers?
\item How efficient are syntax-based log parsers compared to
semantic-based log parsers in terms of runtime?
\item How do semantic-based log parsers perform on unseen log data?
\end{enumerate}
\noindent\textbf{Category 2.} Log Parser Architecture.
\begin{enumerate}[label=\textbf{RQ\arabic{*}.}, start=4]
\item How does two-phase parsing compare with single-phase parsing?
\item Does the order of the log grouping phase and the template
identification phase have any impact on the accuracy?
\end{enumerate}

\subsection{Study Setup}

\subsubsection{Evaluation Metrics}
\label{sec: metrics}

Following Jiang et al.\cite{Loghub2}, we use two categories of metrics, namely
message-level and template-level metrics.  Message-level metrics assess the
parsing of each log message individually, thus favouring templates with high
volume of log messages.  Template-level metrics assess the parsing of the log
templates, therefore eliminating the bias of uneven templates in terms of log
volume.  For both categories of metrics, we choose two metrics: one for
assessing the clustering ability and another for the token classification
ability.


\noindent\textbf{Message-Level Metrics.}
Following existing studies, we utilize two popular message-level metrics, GA and
PA.

\paragraph{Grouping Accuracy (GA)} Proposed by Zhu et al.~\cite{zhu19},
GA assesses the ability to correctly group log messages belonging to the same
template. GA is defined as the ratio of correctly grouped log messages over the
total number of log messages, where a log message is regarded as correctly
grouped if and only if the set of log messages with the same template corresponds
to the same set of log messages in the ground truth.

\paragraph{Parsing Accuracy (PA)} Proposed by Dai et al.\cite{Logram}, PA
assesses the ability to correctly identify the constant parts and variable parts
of each log message. It is defined as the ratio of correctly parsed log messages
over the total number of log messages, where a log message is correctly parsed
if and only if all constant and variable tokens are correctly identified.

\noindent\textbf{Template-Level Metrics.}
Following the recent study by Jiang et al.~\cite{Loghub2}, we use two
template-level metrics, FGA and FTA.

\paragraph{F1-score of Group Accuracy (FGA)} Proposed by Jiang et
al.~\cite{Loghub2}, FGA focuses on the proportion of correctly grouped templates
rather than log messages.  FGA is the harmonic mean of PGA (Precision of Group
Accuracy) and RGA (Recall of Group Accuracy). PGA is defined as the ratio of
correctly grouped log templates over the total number of log templates generated
by the log parser, and RGA is defined as the ratio of correctly grouped log
templates over the total number of log templates in the ground truth.  A log
template is considered correctly grouped if and only if the set of log messages
belonging to this template corresponds to the same set of log messages in the
ground truth.

\paragraph{F1-score of Template Accuracy (FTA)} Proposed by Khan et
al.~\cite{LoghubCorrect}, FTA focuses on the proportion of correctly parsed
templates rather than log messages.  FTA is the harmonic mean of PTA (Precision
of Template Accuracy) and RTA (Recall of Template Accuracy). PTA is defined as
the ratio of correctly parsed log templates over the total number of log
templates generated by the log parser, whereas RTA is defined as the ratio of
correctly parsed log templates over the total number of log templates in the
ground truth.  A log template is considered correctly parsed if and only if all
the log messages belonging to this template are correctly grouped and all the
tokens of the template are the same as those of the ground-truth template.

\subsubsection{Log Parsers}\label{sec: parsers}

For RQ1-RQ3, we choose six state-of-the-art syntax-based log parsers, two for
each of the three categories: AEL and Drain for heustic-based log parsers, SHISO
and LogMine for similarity-based log parsers, LFA and LogCluster for
frequency-based log parsers.  For the semantic-based log parsers, we choose two
state-of-the-art DL-based log parsers UniParser and LogPPT, and two LLM-based
log parsers: LLMParser and LILAC.  8 of these log parsers have previously been
studied by Jiang et al.~\cite{Loghub2}.  We have included LLMParser and LILAC to
incorporate LLM-based log parsing in our study.

For RQ4-5, we study the two-phase log parser Log3T~\cite{Log3T}, which includes
both a semantic-based classifier module for identifying constants and variables
and a syntax-based grouping module for log grouping. By studying the impact of
each of the modules in the effectiveness of log parsing, we aim to understand
the impact of single-phase and two-phase log parsing.

\begin{figure*}[t]
  \centering
  \includegraphics[width=\textwidth]{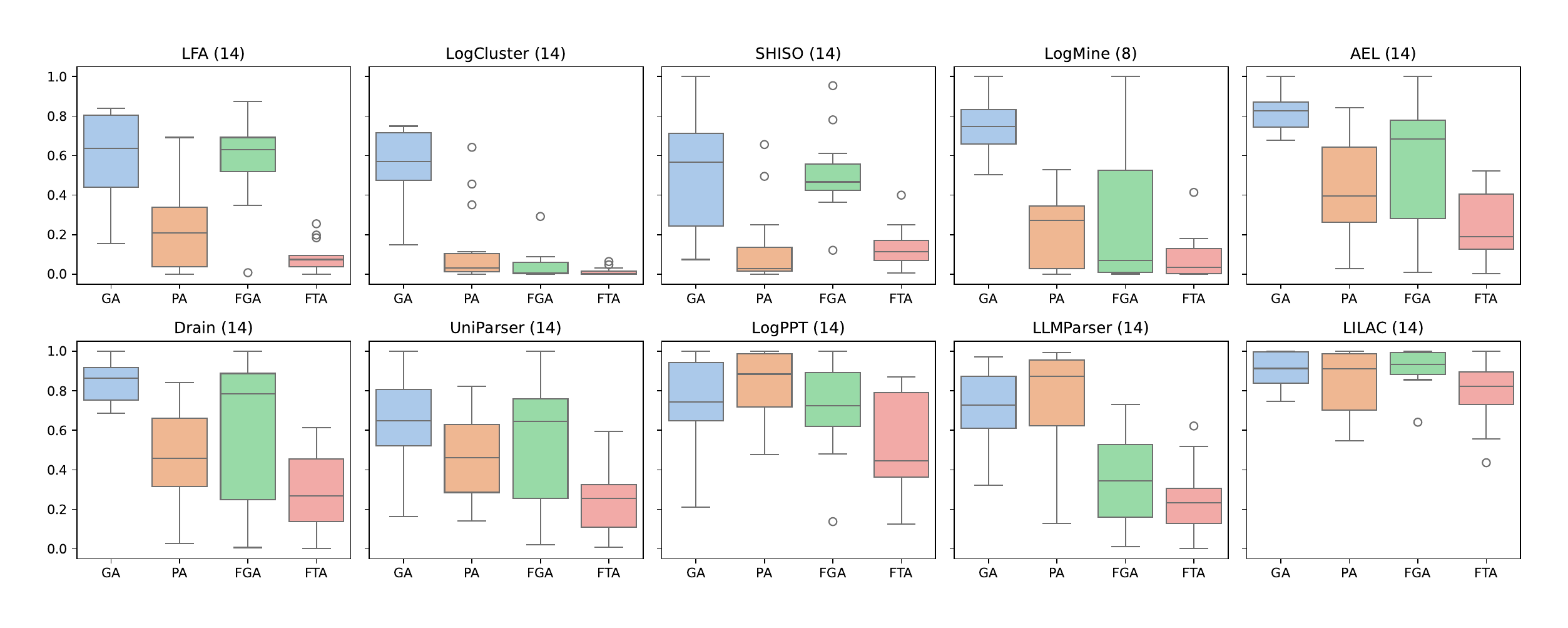}
  \caption{
    Performance evaluation of benchmark syntax-based and semantic-based log parsers
    (number of datasets processed by the parser in parentheses)
  }
  \label{fig: mvgp}
\end{figure*}

\subsubsection{Datasets}\label{sec: loghub2}

For RQ1 - RQ3, we use the Loghub-2 benchmark datasets~\cite{Loghub2}. Loghub-2
contains 14 large-scale log datasets, containing varying sizes of log data
ranging from 21K to more than 16M.


For RQ4 and RQ5, we use the Loghub-2k datasets~\cite{Loghub} instead of Loghub-2
datasets because Log3T fails to parse any of the datasets in Loghub-2 within a
reasonable runtime of 12 hours. Loghub-2k datasets comprise of log data from 16
systems. Each dataset contains 2,000 log messages. Khan et al.~\cite{LoghubCorrect}
detected some issues in the Loghub dataset and published a corrected benchmark.
Following recent research~\cite{LogPPT, LLMParser}, we use the corrected
benchmark dataset.

\subsubsection{Threats to Validity}

The selection of log parsers is limited, as not all existing log parsers are
open-sourced due to industry confidentiality reasons~\cite{SPINE}. Nevertheless,
the selected parsers include state-of-the-art log parsers published at top-tier
conferences and cover all existing categories of technology.

In this study, we have reused the implementations published by Jiang et al. with
Loghub-2~\cite{Loghub2}.  The hyperparameters may not be perfectly tuned for the
new and large dataset since most of the existing log parsers were studied based
on the earlier and smaller Loghub-2k dataset~\cite{Loghub}.

For RQ4 and RQ5, we conduct the experiments only on Log3T, which may bring to
question the applicability of the results in general.  However, we have set the
experiments by analyzing and comparing all possible combinations of both
single-phase and two-phase parsing architectures, thereby eliminating the
concern of general applicability of the study.

\subsection{RQ1:  Performance comparison of syntax- and semantic-based parsers}

In this RQ, we examine the performances of syntax-based log parsers and
semantic-based log parsers.  As discussed in the Study Design section,
syntax-based log parsers employ various pattern mining methods, e.g., token
matching, frequent pattern mining, longest common subsequence, etc., to cluster
log messages so that the log messages of each cluster has the same log
template~\cite{Drain, AEL}. On the other hand, semantic-based log parsers train
a neural network or LLM on sample log data to identify the constant and the
variable tokens~\cite{UniParser, LogPPT}.  We aim to analyze the strengths and
shortcomings of each category of log parsers by conducting a comprehensive
evaluation with four accuracy metrics. The experiments of this RQ are conducted
on the Loghub-2 benchmark dataset~\cite{Loghub2}.

Figure~\ref{fig: mvgp} presents the effectiveness of the log parsers in terms of
the four accuracy metrics. We also present the number of datasets each log
parser can finish processing within 12 hours in parenthesis . According to the
figure, the syntax-based log parsers achieve drastically different GA and PA,
whereas the GA and PA achieved by the semantic-based log parsers are
comparatively closer. Specifically, the syntax-based parsers on average achieve
0.78 and 0.27 GA and PA respectively, thus taking a 51\% hit in performance when
identifying the log template.  On the other hand, semantic-based log parsers
achieve on average 0.75 and 0.72 GA and PA respectively, thus achieving similar
performance in both log grouping and template identification.


Similar characteristics is seen in template-level accuracies.  The syntax-based
log parsers achieve an average of 0.43 and 0.14 FGA and FTA respectively, with a
67\% hit in identifying the template.  On the other hand, the semantic-based log
parsers obtain on average of 0.63 and 0.45 FGA and FTA respectively, with a 28\%
drop in template identification performance. Although the decline is much more
prominent for both categories of log parsers in template-level metrics than in
message-level metrics, the characteristics remain that semantic-based log
parsers achieve comparable performances in terms of both log grouping and
template identification, whereas syntax-based log parsers exhibit a sharp
decline in the latter compared to the former.

For syntax-based log parsers, log grouping is the main phase.  After the
clustering, the template for each log group is identified.  However, from the
stark drop in PA from GA, it is clear that among the logs that are correctly
grouped, the log templates of a vast amount are not correctly identified.
Improving the template identification process of syntax-based log parsers may
drastically improve their PA and FTA.

\rqboxc { \textbf{O1.} Semantic-based approaches are better at identifying log
template of individual log messages whereas syntax-based approaches are
better at grouping the log messages.  The sharp decline in PA from GA calls for
improvement in template identification phase of syntax-based log parsers. }

\subsection{RQ2:  Efficiency comparison of syntax- and semantic-based log
parsers}


\begin{figure}[t]
  \centering
  \begin{subfigure}{0.49\textwidth}
    \includegraphics[width=\linewidth]{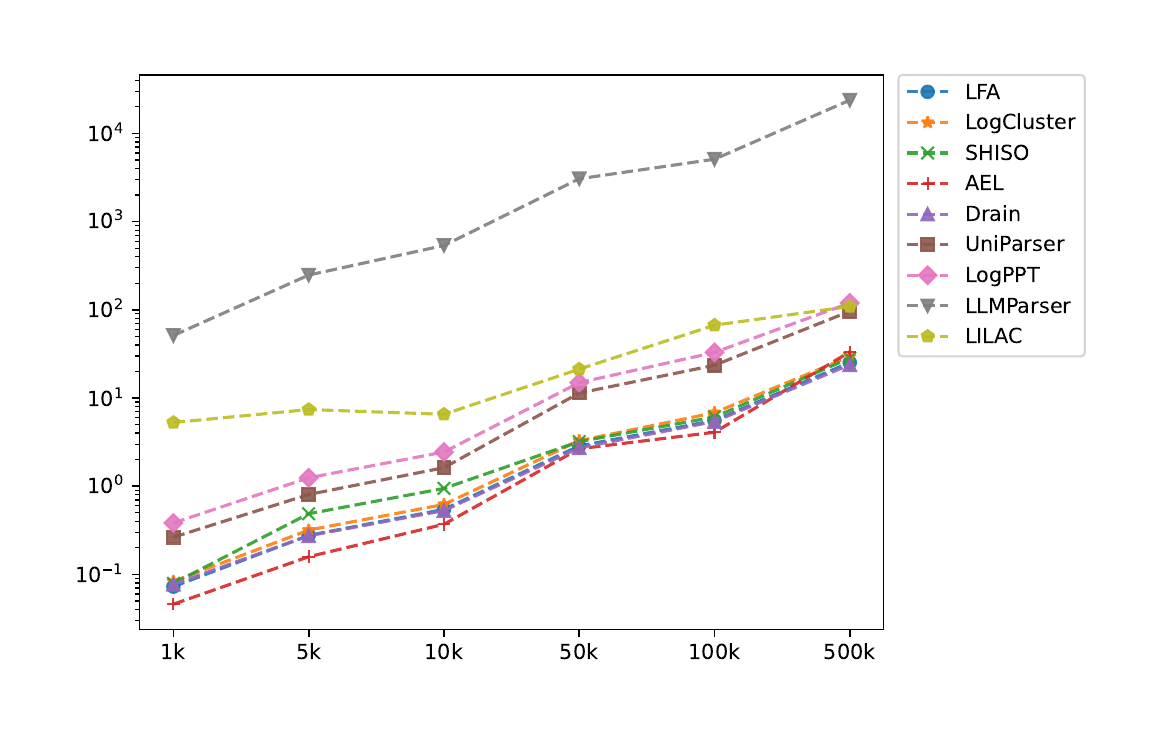}
    \caption{HPC dataset}
  \end{subfigure}
  \begin{subfigure}{0.49\textwidth}
    \includegraphics[width=\linewidth]{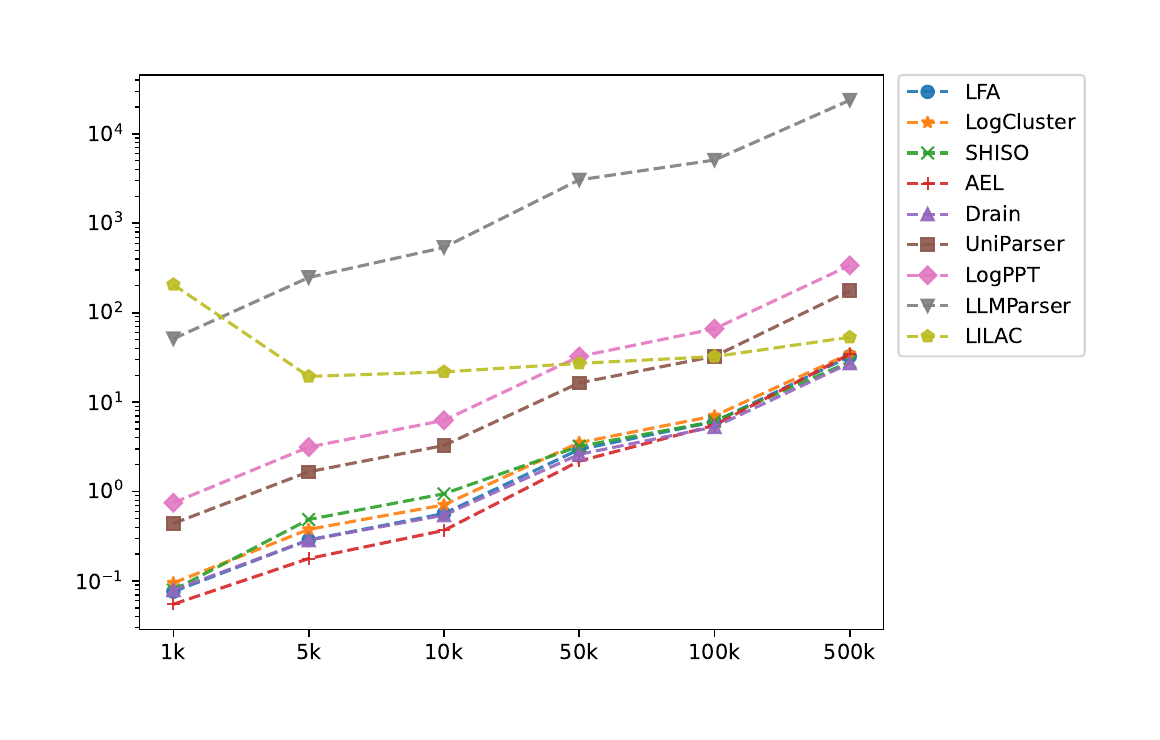}
    \caption{OpenSSH dataset}
  \end{subfigure}
  \caption{Runtime efficiency of benchmark syntax-based and semantic-based log parsers}
  \label{fig: mvge}
\end{figure}

In this RQ, we compare the benchmark syntax- and semantic-based log parsers in
terms of their runtime efficiency, which is a critical metric for large-scale
log parsing. Following LogPPT~\cite{LogPPT}, we evaluate runtime efficiency
using two datasets, HPC and OpenSSH, since they are the largest among the
benchmark datasets.

Figure~\ref{fig: mvge} presents the runtime of the three best-performing
benchmark log parsers (AEL, Spell, and Drain) on HPC and OpenSSH datasets in a
logarithmic scale.  The figure shows that syntax-based log parsers require
significantly smaller execution time compared to the inference time of the
semantic-based log parsers.  Specifically, the LLM-based log parser, LLMParser,
requires $10^2 = 100$ times more computation time compared to the DL-based log
parsers.  LLM-based log parsers are clearly inadequate for the demands of
large-scale log parsing.  The syntax-based log parsers, on the other hand,
require 10 times less computation time than DL-based log parsers and 1K times
less computation time than the LLM-based log parser.  Hence, in terms of
efficiency, syntax-based log parsers are the most suitable for large-scale
parsing.

\rqboxc {
  \textbf{O2.} Syntax-based parsers are significantly faster, requiring
  10x-1,000x less time than semantic-based parsers. The runtime efficiency makes
  syntax-based parsers more suitable for scalable large-scale use.
}

\subsection{RQ3:  Performance of semantic-based parsers on unseen log data}

\begin{figure}[t]
  \centering
  \includegraphics[width=0.80\columnwidth]{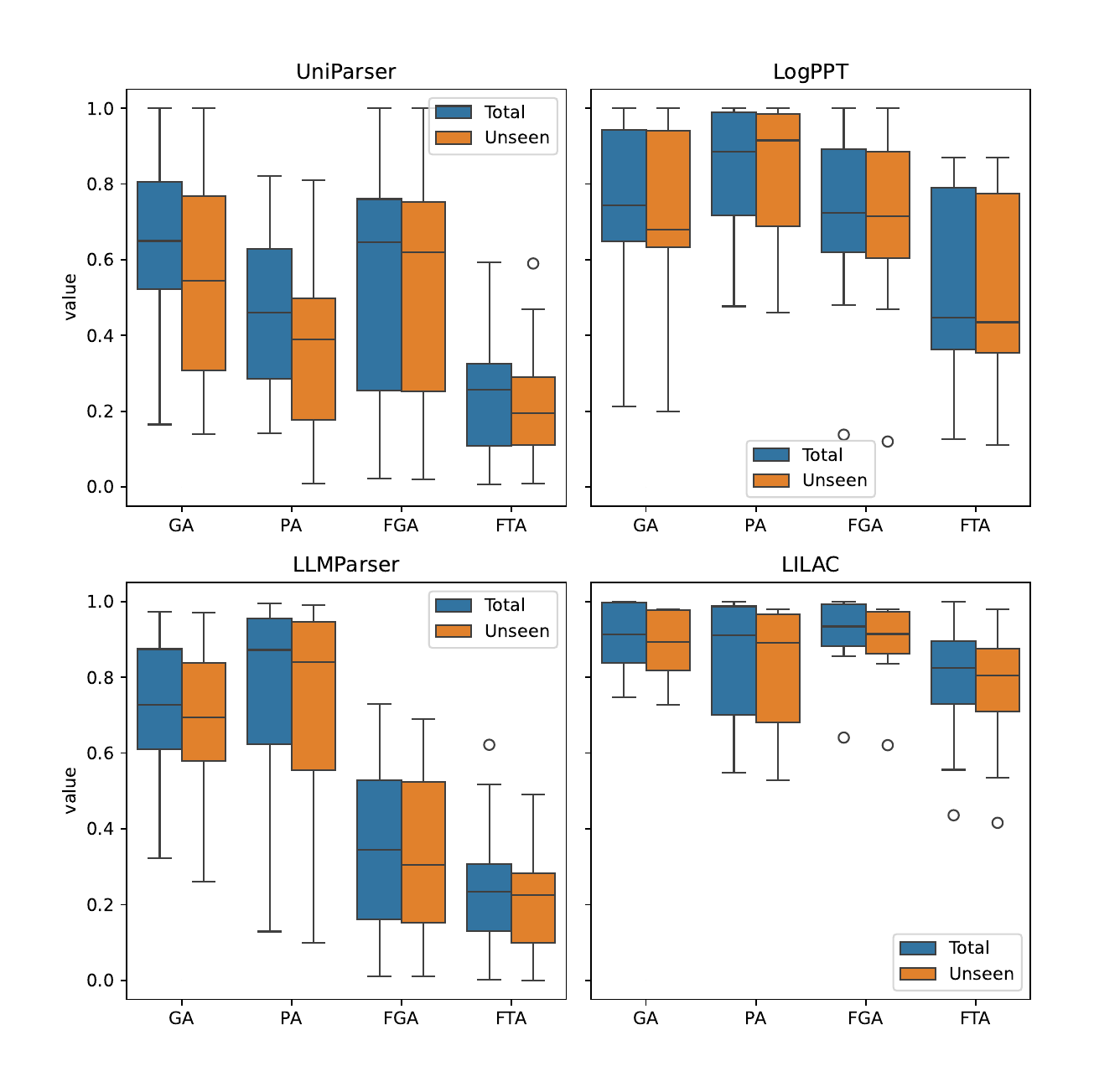}
  \caption{Performance of semantic-based log parser on unseen log data}
  \label{fig: unseen}
\end{figure}

Semantic-based log parsers require training on a set of sample log data.  The
log parser trains on this sample log data and learns the semantic features of
the whole dataset.  But prior studies have shown that semantic-based log parsers
face challenges generalizing over unseen log data, particularly when the
training set is limited~\cite{Loghub2}.

In this RQ, we study the difference in performance of semantic-based log parsers
when evaluated on unseen log data compared to when evaluated on log data
including the training samples.



\begin{table}[t]
  \centering
\caption{Comparison of single-phase and two-phase log parsing}\label{tab: 1v2}
  \scriptsize
\begin{tabular}{lrrrrrrrrr}\toprule
&\multicolumn{2}{c}{\textbf{GA}} &\multicolumn{2}{c}{\textbf{PA}} &\multicolumn{2}{c}{\textbf{FGA}} &\multicolumn{2}{c}{\textbf{FTA}} \\\cmidrule{2-9}
  \textbf{Dataset} &\multicolumn{1}{c}{\textbf{Single}} &\multicolumn{1}{c}{\textbf{Two}} &\multicolumn{1}{c}{\textbf{Single}} &\multicolumn{1}{c}{\textbf{Two}} &\multicolumn{1}{c}{\textbf{Single}} &\multicolumn{1}{c}{\textbf{Two}} &\multicolumn{1}{c}{\textbf{Single}} &\multicolumn{1}{c}{\textbf{Two}} \\
  &\textbf{Phase} &\textbf{Phase} &\textbf{Phase} &\textbf{Phase} &\textbf{Phase} &\textbf{Phase} &\textbf{Phase} &\textbf{Phase} \\\midrule
\textbf{Apache} &\textbf{0.86} &0.84 &0.16 &\textbf{0.17} &0.72 &\textbf{0.85} &0.15 &\textbf{0.31} \\
\textbf{Android} &\textbf{1.00} &\textbf{1.00} &\textbf{0.98} &0.69 &\textbf{1.00} &\textbf{1.00} &\textbf{0.67} &0.50 \\
\textbf{BGL} &0.88 &\textbf{0.98} &0.33 &\textbf{0.35} &0.66 &\textbf{0.89} &0.15 &\textbf{0.24} \\
\textbf{HDFS} &0.81 &\textbf{1.00} &0.50 &\textbf{0.90} &0.05 &\textbf{1.00} &0.04 &\textbf{0.71} \\
\textbf{HPC} &0.90 &\textbf{0.90} &0.63 &\textbf{0.66} &0.26 &\textbf{0.78} &0.14 &\textbf{0.47} \\
\textbf{Hadoop} &\textbf{0.98} &0.95 &0.35 &\textbf{0.37} &\textbf{0.88} &0.87 &0.28 &\textbf{0.50} \\
\textbf{HealthApp} &1.00 &\textbf{1.00} &0.18 &\textbf{0.18} &0.98 &\textbf{1.00} &0.35 &\textbf{0.37} \\
\textbf{Linux} &0.29 &\textbf{0.83} &\textbf{0.16} &0.11 &\textbf{0.81} &0.79 &0.28 &\textbf{0.49} \\
\textbf{Mac} &0.71 &\textbf{0.90} &0.28 &\textbf{0.29} &0.72 &\textbf{0.82} &0.20 &\textbf{0.27} \\
\textbf{OpenSSH} &0.39 &\textbf{0.82} &0.24 &\textbf{0.30} &0.18 &\textbf{0.31} &0.03 &\textbf{0.08} \\
\textbf{OpenStack} &0.25 &\textbf{0.48} &\textbf{0.11} &0.10 &0.21 &\textbf{0.79} &0.09 &\textbf{0.13} \\
\textbf{Proxifier} &0.00 &\textbf{1.00} &\textbf{0.00} &\textbf{0.00} &0.00 &\textbf{1.00} &\textbf{0.00} &\textbf{0.00} \\
\textbf{Spark} &0.74 &\textbf{0.92} &0.33 &\textbf{0.40} &0.24 &\textbf{0.87} &0.09 &\textbf{0.55} \\
\textbf{Thunderbird} &0.92 &\textbf{0.95} &0.03 &\textbf{0.04} &0.68 &\textbf{0.80} &0.24 &\textbf{0.39} \\
\textbf{Windows} &0.71 &\textbf{0.99} &0.14 &\textbf{0.16} &0.73 &\textbf{0.82} &0.15 &\textbf{0.42} \\
\textbf{Zookeeper} &0.97 &\textbf{0.99} &0.50 &\textbf{0.50} &0.86 &\textbf{0.89} &0.40 &\textbf{0.48} \\\midrule
\textbf{Average} &0.71 &\textbf{0.91} &0.31 &\textbf{0.33} &0.56 &\textbf{0.84} &0.20 &\textbf{0.37} \\
\bottomrule
\end{tabular}
\end{table}

Figure~\ref{fig: unseen} presents the performances of the semantic-based log
parsers on the whole benchmark dataset and on only the unseen log data.  All
three benchmark semantic-based log parsers demonstrate a drop in performance
when evaluated only on unseen log data.  While LogPPT shows a reasonable drop
ranging from 1\% to 3\%, both UniParser and LLMParser suffer a drop ranging
from 12\% to 23\% and from 4\% to 15\% respectively.  The results indicate the
lack of generalization ability of semantic-based log parsers across unseen log
data.

\rqboxc {
  \textbf{O3.} All three semantic-based log parsers show performance drops
  across all metrics when evaluated only on unseen log data. This is indicative
  of semantic-based log parsers lacking in generalization ability.
}



\subsection{RQ4:  Comparison of single-phase and two-phase parsing}

In this RQ, we compare single-phase and two-phase parsing architectures. In
single-phase log parsing, only one log parsing technique is used, whereas in
two-phase log parsing, both syntax-based log grouping and semantic-based token
classification is used.  For example, Log3T is comprised of two modules: (1) a
BERT-based classifier that classifies the top $k$ tokens in a log as most likely
to be constants and (2) a grouping module that groups logs with the same
constant tokens in the same order.  To study the impact of two-phase parsing in
comparison to single-phase parsing, we modify Log3T and remove the grouping
module, effectively transforming it into a single-phase log parser.

Table~\ref{tab: 1v2} presents the performances obtained by single-phase parsing
and two-phase parsing.  Two-phase parsing achieves better accuracy across all
four metrics, with the performance improving from 27\% in GA to 81\% in FTA. The
improvement in PA, however, is only 6.17\%, which is statistically
insignificant. The results suggest that the two-phase architecture substantially
enhances a log parser's ability to correctly group log messages belonging to the
same template, although it has limited impact on the message-level accuracy of
the log templates.

\rqboxc{ \textbf{O4.} Two-phase log parsing achieves better accuracy across all
four metrics, with the improvements being more significant in grouping-related
metrics.}

\subsection{RQ5:  Impact of log grouping before token classification}

\begin{table}[t]\centering
\caption{Impact of the order of Log Grouping and Token Classification phases in two-phase log parsing}
\label{tab: log3t2}
  \scriptsize
\begin{tabular}{lrrrrrrrrr}\toprule
&\multicolumn{2}{c}{\textbf{GA}} &\multicolumn{2}{c}{\textbf{PA}} &\multicolumn{2}{c}{\textbf{FGA}} &\multicolumn{2}{c}{\textbf{FTA}} \\\cmidrule{2-9}
\textbf{Dataset} &\textbf{Log3T} &\textbf{Log3T}' &\textbf{Log3T} &\textbf{Log3T}' &\textbf{Log3T} &\textbf{Log3T}' &\textbf{Log3T} &\textbf{Log3T}' \\\midrule
\textbf{Apache} &\textbf{0.84} &0.81 &\textbf{0.17} &0.13 &\textbf{0.85} &0.82 &\textbf{0.31} &0.29 \\
\textbf{Android} &\textbf{1.00} &\textbf{1.00} &\textbf{0.69} &\textbf{0.69} &\textbf{1.00} &\textbf{1.00} &\textbf{0.50} &\textbf{0.50} \\
\textbf{BGL} &0.98 &\textbf{0.98} &\textbf{0.35} &\textbf{0.35} &0.89 &\textbf{0.90} &0.24 &\textbf{0.24} \\
\textbf{HDFS} &\textbf{1.00} &\textbf{1.00} &\textbf{0.90} &\textbf{0.90} &\textbf{1.00} &\textbf{1.00} &\textbf{0.71} &\textbf{0.71} \\
\textbf{HPC} &0.90 &\textbf{0.91} &\textbf{0.66} &\textbf{0.66} &0.78 &\textbf{0.83} &0.47 &\textbf{0.49} \\
\textbf{Hadoop} &\textbf{0.95} &\textbf{0.95} &\textbf{0.37} &\textbf{0.37} &\textbf{0.87} &\textbf{0.87} &\textbf{0.50} &\textbf{0.50} \\
\textbf{HealthApp} &\textbf{1.00} &\textbf{1.00} &\textbf{0.18} &\textbf{0.18} &\textbf{1.00} &\textbf{1.00} &\textbf{0.37} &\textbf{0.37} \\
\textbf{Linux} &\textbf{0.83} &0.77 &\textbf{0.11} &0.10 &\textbf{0.79} &0.75 &\textbf{0.49} &0.46 \\
\textbf{Mac} &\textbf{0.90} &\textbf{0.90} &\textbf{0.29} &\textbf{0.29} &0.82 &\textbf{0.83} &\textbf{0.27} &\textbf{0.27} \\
\textbf{OpenSSH} &\textbf{0.82} &\textbf{0.82} &\textbf{0.30} &\textbf{0.30} &\textbf{0.31} &\textbf{0.31} &\textbf{0.08} &\textbf{0.08} \\
\textbf{OpenStack} &0.48 &\textbf{0.95} &\textbf{0.10} &\textbf{0.10} &0.79 &\textbf{0.85} &0.13 &\textbf{0.14} \\
\textbf{Proxifier} &0.52 &\textbf{1.00} &\textbf{0.00} &\textbf{0.00} &0.80 &\textbf{1.00} &\textbf{0.00} &\textbf{0.00} \\
\textbf{Spark} &\textbf{0.92} &0.92 &\textbf{0.40} &0.40 &\textbf{0.87} &0.82 &\textbf{0.55} &0.48 \\
\textbf{Thunderbird} &0.95 &\textbf{0.95} &0.04 &\textbf{0.04} &0.80 &\textbf{0.81} &0.39 &\textbf{0.39} \\
\textbf{Windows} &\textbf{0.99} &\textbf{0.99} &\textbf{0.16} &\textbf{0.16} &\textbf{0.82} &\textbf{0.82} &\textbf{0.42} &\textbf{0.42} \\
\textbf{Zookeeper} &\textbf{0.99} &0.99 &\textbf{0.50} &0.50 &\textbf{0.89} &0.86 &\textbf{0.48} &0.45 \\\midrule
\textbf{Average} &0.88 &\textbf{0.93} &\textbf{0.33} &0.33 &0.83 &\textbf{0.84} &\textbf{0.37} &0.36 \\
\bottomrule
\end{tabular}
\end{table}

In this RQ, we modify Log3T, hereby referred to as Log3T', by moving the
grouping module before the classification module.  In the original Log3T, at
first the tokens of the log message is given a likelihood of being a constant or
a variable.  The top $k$ tokens most likely to be constants are then used to
cluster the log message. In the modification for this RQ, Log3T' first tries to
match the log message with existing log templates, i.e., templates of the log
messages that have already been parsed, and if no match is found, Log3T' uses
the classifier probabilities to find a match. If no match is found this time
either, a new log group is created with the current log message as its sole
member (this last step is unchanged from Log3T).

Table~\ref{tab: log3t2} presents the performances obtained by Log3T and Log3T'.
Log3T' achieves 5.4\% higher GA and 1.2\% higher FGA than Log3T.  However,
Log3T' suffers a loss of 0.2\% and 0.7\% in PA and FTA respectively compared to
Log3T. So, prioritizing the template identification phase before the grouping phase
provides a slightly better performance in terms of template-related metrics,
whereas prioritizing the grouping phase before the template identification phase
provides a significant performance boost in terms of grouping-related metrics.

\rqboxc{
  \textbf{O5.} Placing the log grouping phase before the template identification phase in a two-phase
  log parser provides a significant performance boost in grouping-related metrics
  but incurs a slight performance loss in template-related metrics.
}

\section{SynLog+}

\begin{table*}[t]\centering
\caption{Takeaways from the empirical study}
\label{tab: disc}
\small
\begin{tabularx}{1\textwidth}{lXXX}
  \toprule
  \textbf{RQ} &\textbf{Observation} &\textbf{SynLog+ Design Rationale} &\textbf{SynLog+ Modules}
  \\\midrule
  RQ4 &Two-phase parsing performs better than single-phase parsing across all
  four evaluation metrics (O4) &The tasks of log grouping and template identification
  should be modularized &The Log Grouping module is isolated from the other
  modules
  \\\midrule
  RQ5 &Log grouping before token classification results in significantly higher
  grouping accuracy at slight cost of template accuracy (O5) &Grouping should be
  performed prior to identifying the constants and variables &The Log Grouping
  module precedes the other modules
  \\\midrule
  RQ1 &Syntax-based log parsers are better at grouping similar log messages
  while semantic-based log parsers are better at identifying the log templates
  (O1) &Syntax-based log parsers should be used for clustering the logs into log
  groups &Existing syntax-based log parsers are treated as Log Grouping module
  \\\midrule
  RQ2 &Semantic-based log parsers require exponentially more runtime than
  syntax-based log parsers (O2) &Semantic-based techniques should be avoided for
  large-scale log parsing &SynLog+ does not incorporate semantic-based techniques
  in its modules
  \\\midrule
  RQ3 &Semantic-based log parsers lack in generalization ability in untrained
  log data (O3) &Semantic-based approaches should be avoided to ensure generalization
  &SynLog+ does not incorporate semantic-based techniques in its modules
  \\\bottomrule
\end{tabularx}
\end{table*}

\begin{figure}[t] \centering
  \includegraphics[width=0.9\columnwidth]{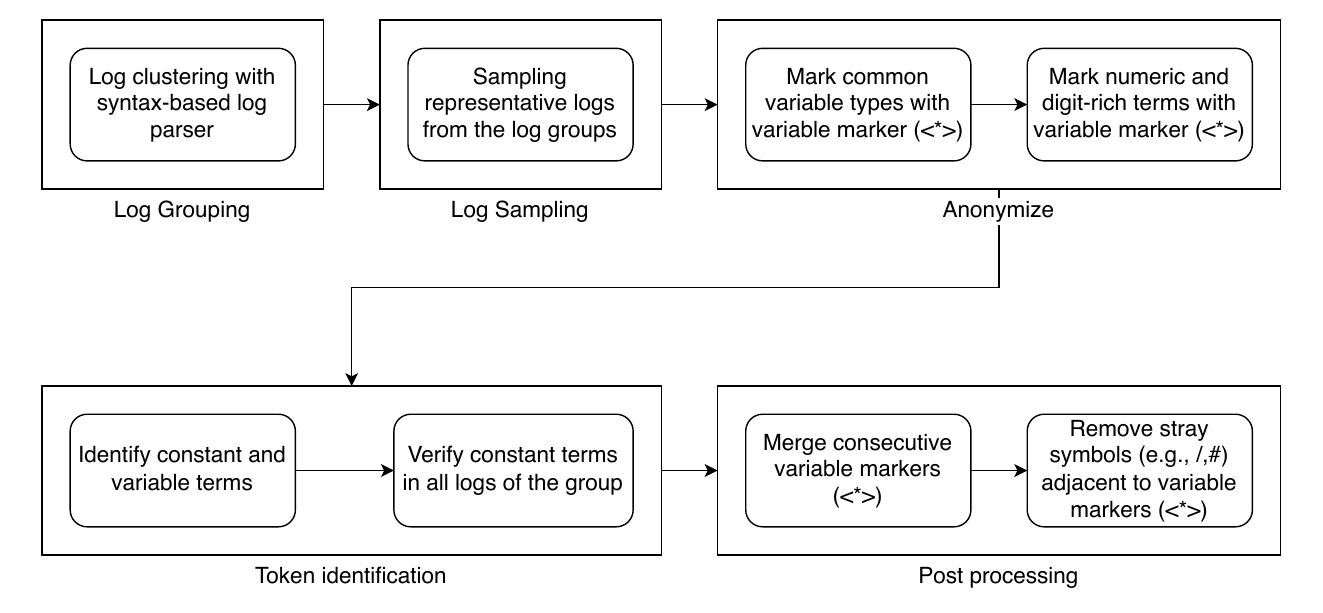}
  \caption{SynLog+ workflow}
  \label{fig: pipe}
\end{figure}

Table~\ref{tab: disc} summarizes the key insights from our empirical study,
which influenced the design decisions of our technique, SynLog+.
Figure~\ref{fig: pipe} displays the modules and workflow of SynLog+.

The study revealed that two-phase parsing outperforms single-phase parsing
across all four evaluation metrics (O4) and performing log grouping before
template identification improves the grouping accuracy (O5). In response,
SynLog+ is designed to have the Log Grouping module isolated and preceding
the rest of the pipeline.

The study also confirms that syntax-based log parsers are more effective for
grouping similar log messages, while semantic-based log parsers perform better
at template identification (O1). However, the latter comes at a substantial
cost, both in terms of runtime efficiency (O2) and generalization across unseen
log formats (O3). Given these limitations, SynLog+ avoids incorporating
semantic-based models in favor of syntax-based techniques.


\subsection{Log grouping}

In SynLog+, the clustering of similar logs is done by an existing syntax-based
log parser.  The Log Grouping module is a syntax-based log parser, which
clusters similar logs in log groups.  The subsequent modules of SynLog+ then
identifies the log template of each log group by identifying the constant and
variable terms in the constituent logs of the log groups.

\subsection{Log sampling}

The overall objective of SynLog+ is to identify variables among the logs in a
log group.  For this, we have to iterate over all the constituent log messages.
But this would be severely time consuming.  So, instead of iterating over all
the logs in a log group, we randomly choose two log messages as representative
logs from the set of unique logs in a log group and move on to the next steps.

\subsection{Anonymize}

Following AEL~\cite{AEL}, we begin with anonymization, i.e., identifying likely
variable tokens in the logs.  Anonymization boosts template accuracy by marking
variables with heuristics before employing any syntax- or semantic-based methods.
To anonymize variable tokens, we replace them with the variable marker
``\verb!<*>!''.  In this step, we design the anonymize step based on the
following observations.

\textbf{Regex for common variable patterns}.  Although initially most of the log
parsers made use of regex patterns to preprocess the log messages~\cite{Drain},
recently researchers~\cite{LogPPT} have argued that using regex patterns are not
feasible in real life scenarios because of two issues: 1) they are specific to
individual log datasets, and 2) they need to be updated with the update of the
software system. Although these two arguments are valid, they concern only a
subset of regex patterns.  Some regex patterns are indeed specific to a dataset
and require regular updates to match the evolution of the software (e.g.,
``blk\_2345'' in BGL), but many variable terms consist of common patterns across
any and all log dataset (e.g., IP addresses, MAC addresses, file paths, etc.).
Since the issues argued against the use of regex patterns in log parsing do not
apply to these common patterns, we used these regex patterns to anonymize the
log messages.

\textbf{Numbers are variables}.  Most log parsers, even the semantic-based ones,
regard numbers as variables.  Some log parsers only consider pure numbers (i.e.,
real or floating numbers in decimal or hexadecimal) as variables, whereas some
log parsers consider terms with $p\%$ digit characters as variables. In
SynLog+, we consider those terms as variables which are either pure
numbers or contain more digits than non-digit characters.

\subsubsection{Anonymizing common variable patterns}

\begin{table}[tp]\centering
  \caption{Regex patterns for common variables}\label{tab: reganon}
    \scriptsize
  \begin{tabular}{llrr}\toprule
    \textbf{Variable Category} &\textbf{Variable Type} &\textbf{Regex Pattern} \\\midrule
    \multirow{8}{*}{Location Identifier} &IP address &
    \verb!(?:[-0-9a-zA-Z]+\.){2,}[-0-9a-zA-Z]+(?::?:\d+)?! \\
    \cmidrule{2-3}
    &MAC address &
    \verb!([A-Fa-f0-9]{2}:){5,11}[A-Fa-f0-9]{2}! \\
    \cmidrule{2-3}
    &Email address &
    \verb![0-9a-zA-Z]+@([0-9a-zA-Z]+\.)+[0-9a-zA-Z]+! \\
    \cmidrule{2-3}
    &\multirow{2}{*}{Unix path}
    &\verb!(\/[\d+\w+\-_\.\\\#\$]*[\/\.]! \\
    & &\verb![\d+\w+\-_\.\\\#\$\/*]*)+(\sHTTPS?\/\d\.\d)?! \\
    \cmidrule{2-3}
    &\multirow{2}{*}{Windows path}
    &\verb!([a-zA-Z]\:[\/\\][\d+\w+\-_\.\\\#\$]*! \\
    & &\verb!([\/\\\.][\d+\w+\-_\.\\\#\$\\\/*]*)?)! \\
    \midrule
    \multirow{2}{*}{Time/Duration} &Date \& Time &
    \verb!(\d{1,4}(-|/)\d{1,2}(-|/)\d{1,4})! \\
    \cmidrule{2-3}
    &Duration &
    \verb![+-]?(\d+s(\d+\s?ms)?|\d+\s?ms)! \\
    \midrule
    Computing Resource &Memory &
    \verb!(\d+(\.\d+)?)\s?[kmgKMG][bB]?((\/s)|(ytes))?! \\
    \bottomrule
  \end{tabular}
\end{table}

In this step, we match for common variable patterns with regex matching. Li et
al. conducted a study exploring the characteristics of dynamic variables in log
messages~\cite{VarAware}.  Among the 9 categories of variables in their study, 5
are numbers, which we have already anonymized in the previous step. Among the
other 4 categories, Object Name (OBN) do not follow any universal pattern.  The
other three categories include (1) Location Indicator, (2) Time/Duration of an
Action, and (3) Computing Resources.  Location Indicator can be path
information, a URI, IP, or MAC address.  Time/Duration of an Action shows the
execution time or duration of an action.  It can either be a precise date and
time information (e.g., ``Sat Jun 18 02:08:10 2005'') or a duration of time in
seconds, minutes, or hours (e.g., ``Scheduled snapshot period at 10ms'').
Computing Resources include the amount of memory (e.g., 126MB) or clock cycle
(e.g., 1.3 MHz) consumed by a process.  We design regex patterns, presented in
Table~\ref{tab: reganon}, to capture these three categories of variables in the
log messages.

\subsubsection{Anonymizing numbers}

In this step, we anonymize the terms which satisfy one of the two following
criteria.

\textbf{Pure number}.  The term is a pure number, i.e., if it is a real
  number in decimal or hexadecimal systems.

\textbf{Rich in digits}. The term consists mostly of digit characters
  than of non-digit characters.

\subsection{Identifying constants and variables}

After the anonymize step, we iterate over the terms of the two representative
log message for each log group and identify those terms as constants which are
present in both log messages \textit{in the same order} and those terms as
variables which are not present in both log messages.  The variable terms
identified in this manner are indeed variable terms in the log template.
However, the constant terms thus far identified may not be constant terms in the
log template, since they may be absent in the other logs in the log groups.
Thus, after identifying the variable and constant terms in the two
representative logs, we iterate over the constant terms and search for them in
all the logs.  If absent in any of the logs, we mark the constant term as
variable instead.

\subsection{Post processing}

In this step, multiple consecutive variable markers are substituted with a
single variable marker (\verb|<*>|).  Moreover, due to imprecise tokenization,
the variable markers in the log template may have stray symbols, e.g., slashes
or pounds, before or after it.  In this step, we remove these stray characters.
In other words, we make these a part of the variable terms.

\section{Evaluation of SynLog+}


We answer the following research
questions (RQs).
\begin{enumerate}[label=\textbf{RQ\arabic{*}.}, start=6, leftmargin=3em]

\item How much performance boost is achieved by SynLog+ when
coupled with baseline log parsers?

\item How efficient is SynLog+ compared to the baseline log
parsers?

\item What impact does the number of representative logs chosen have on the
  performance?

\item How generalizable are the regex patterns used by SynLog+?

\end{enumerate}
\subsection{Study Setup}

\subsubsection{Datasets}

For the evaluation, we use the Loghub-2 benchmark~\cite{Loghub2} as described in
Section~\ref{sec: loghub2}.

\subsubsection{Evaluation Metrics}

We use the same set of message-level and template-level evaluation metrics
discussed in Section~\ref{sec: metrics}.

\subsubsection{Log Parsers}

We choose the same 12 state-of-the-art log parsers from Section~\ref{sec:
parsers} as baselines.


\subsubsection{Threats to Validity}

SynLog+ relies on regex pattern matching to anonymize common variables.  We 
followed Li et al.~\cite{VarAware} to identify common variable
types.  The regex patterns utilized in this work may not detect all the variable
types (see RQ8).  Further work is required to catalog all regex patterns.

\subsection{RQ6: Effectiveness Evaluation}

\begin{table*}[tp]\centering
\caption{Improvement in GA}\label{tab: ga}
\begin{adjustbox}{width=1\textwidth}
\scriptsize
\begin{tabular}{lrrrrrrrrrrrrrrrrrrrrr}\toprule
  &\multicolumn{2}{c}{\textbf{LFA}}
  &\multicolumn{2}{c}{\textbf{LogCluster}}
  &\multicolumn{2}{c}{\textbf{SHISO}}
  &\multicolumn{2}{c}{\textbf{LogMine}}
  &\multicolumn{2}{c}{\textbf{AEL}}
  &\multicolumn{2}{c}{\textbf{Drain}}
  &\multicolumn{2}{c}{\textbf{UniParser}}
  &\multicolumn{2}{c}{\textbf{LogPPT}}
  &\multicolumn{2}{c}{\textbf{LLMParser}}
  &\multicolumn{2}{c}{\textbf{LILAC}}\\
    \cmidrule{2-21}
      &\textbf{\xmark} &\textbf{\cmark} &\textbf{\xmark} &\textbf{\cmark} &\textbf{\xmark} &\textbf{\cmark}
      &\textbf{\xmark} &\textbf{\cmark} &\textbf{\xmark} &\textbf{\cmark} &\textbf{\xmark} &\textbf{\cmark}
      &\textbf{\xmark} &\textbf{\cmark} &\textbf{\xmark} &\textbf{\cmark} &\textbf{\xmark} &\textbf{\cmark}
      &\textbf{\xmark} &\textbf{\cmark} \\\toprule
\textbf{Proxifier} &\textbf{0.35} &\textbf{0.35} &0.66 &\textbf{0.67} &0.69 &\textbf{0.98} &0.50 &\textbf{0.52} &\textbf{0.83} &\textbf{0.83} &\textbf{0.69} &\textbf{0.69} &\textbf{0.51} &\textbf{0.51} &\textbf{0.99} &\textbf{0.99} &\textbf{0.96} &\textbf{0.96} &\textbf{1.00} &\textbf{1.00} \\
\textbf{Linux} &0.23 &\textbf{0.69} &0.60 &\textbf{0.65} &0.07 &\textbf{0.52} &0.74 &\textbf{0.76} &0.68 &\textbf{0.70} &0.69 &\textbf{0.71} &0.20 &\textbf{0.68} &0.21 &\textbf{0.22} &0.50 &\textbf{0.96} &\textbf{0.83} &0.83 \\
\textbf{Apache} &\textbf{0.81} &\textbf{0.81} &0.55 &\textbf{0.99} &0.57 &\textbf{0.59} &\textbf{1.00} &\textbf{1.00} &\textbf{1.00} &\textbf{1.00} &\textbf{1.00} &\textbf{1.00} &0.17 &\textbf{0.59} &0.80 &\textbf{0.99} &0.88 &\textbf{1.00} &\textbf{1.00} &\textbf{1.00} \\
\textbf{Zookeeper} &0.84 &\textbf{0.98} &0.74 &\textbf{0.99} &0.82 &\textbf{0.96} &0.70 &\textbf{0.99} &1.00 &\textbf{1.00} &\textbf{0.99} &\textbf{0.99} &0.99 &\textbf{1.00} &0.98 &\textbf{0.99} &0.89 &\textbf{1.00} &\textbf{0.99} &\textbf{0.99} \\
\textbf{Hadoop} &0.83 &\textbf{0.91} &0.48 &\textbf{0.90} &0.72 &\textbf{0.77} &0.83 &\textbf{0.91} &0.82 &\textbf{0.94} &0.92 &\textbf{0.94} &0.59 &\textbf{0.94} &0.68 &\textbf{0.93} &0.87 &\textbf{0.97} &\textbf{0.92} &\textbf{0.92} \\
\textbf{HealthApp} &0.80 &\textbf{0.94} &0.73 &\textbf{0.94} &0.08 &\textbf{0.37} &0.55 &\textbf{0.68} &0.73 &\textbf{0.97} &0.86 &\textbf{0.99} &0.59 &\textbf{0.87} &\textbf{1.00} &\textbf{1.00} &0.74 &\textbf{1.00} &\textbf{0.99} &\textbf{0.99} \\
\textbf{OpenStack} &\textbf{0.67} &\textbf{0.67} &0.69 &\textbf{0.86} &0.81 &\textbf{0.89} &0.76 &\textbf{0.98} &\textbf{0.74} &\textbf{0.74} &0.75 &\textbf{0.82} &0.56 &\textbf{1.00} &\textbf{1.00} &\textbf{1.00} &0.74 &\textbf{1.00} &\textbf{1.00} &\textbf{1.00} \\
\textbf{HPC} &\textbf{0.73} &\textbf{0.73} &0.73 &\textbf{0.73} &0.08 &\textbf{0.08} &\multicolumn{1}{c}{\textemdash} &\multicolumn{1}{c}{\textemdash} &0.75 &\textbf{0.80} &0.79 &\textbf{0.84} &\textbf{0.81} &\textbf{0.81} &0.76 &\textbf{0.84} &0.64 &\textbf{0.70} &\textbf{0.87} &\textbf{0.87} \\
\textbf{Mac} &0.59 &\textbf{0.70} &0.46 &\textbf{0.80} &0.61 &\textbf{0.79} &0.85 &\textbf{0.88} &0.80 &\textbf{0.80} &0.76 &\textbf{0.90} &0.85 &\textbf{0.90} &0.58 &\textbf{0.82} &0.60 &\textbf{0.85} &0.82 &\textbf{0.91} \\
\textbf{OpenSSH} &0.16 &\textbf{0.49} &0.22 &\textbf{0.50} &0.40 &\textbf{0.70} &\multicolumn{1}{c}{\textemdash} &\multicolumn{1}{c}{\textemdash} &\textbf{0.71} &\textbf{0.71} &\textbf{0.71} &\textbf{0.71} &0.28 &\textbf{0.34} &0.28 &\textbf{0.39} &0.32 &\textbf{0.60} &\textbf{0.75} &\textbf{0.75} \\
\textbf{HDFS} &0.81 &\textbf{0.81} &0.49 &\textbf{0.76} &\textbf{1.00} &\textbf{1.00} &\multicolumn{1}{c}{\textemdash} &\multicolumn{1}{c}{\textemdash} &\textbf{1.00} &\textbf{1.00} &\textbf{1.00} &\textbf{1.00} &\textbf{1.00} &\textbf{1.00} &0.82 &\textbf{1.00} &0.97 &\textbf{1.00} &\textbf{1.00} &\textbf{1.00} \\
\textbf{Spark} &0.50 &\textbf{0.84} &0.15 &\textbf{0.76} &0.19 &\textbf{0.46} &\multicolumn{1}{c}{\textemdash} &\multicolumn{1}{c}{\textemdash} &0.83 &\textbf{0.87} &0.87 &\textbf{0.87} &0.77 &\textbf{0.77} &0.69 &\textbf{0.99} &0.72 &\textbf{0.98} &0.79 &\textbf{0.79} \\
\textbf{Thunderbird} &0.61 &\textbf{0.72} &0.47 &\textbf{0.62} &0.50 &\textbf{0.57} &\multicolumn{1}{c}{\textemdash} &\multicolumn{1}{c}{\textemdash} &0.83 &\textbf{0.86} &\textbf{0.88} &0.87 &0.79 &\textbf{0.80} &0.64 &\textbf{0.70} &0.67 &\textbf{0.73} &\textbf{0.90} &0.89 \\
\textbf{BGL} &0.42 &\textbf{0.43} &0.75 &\textbf{0.88} &0.56 &\textbf{0.65} &\multicolumn{1}{c}{\textemdash} &\multicolumn{1}{c}{\textemdash} &0.88 &\textbf{0.89} &0.91 &\textbf{0.91} &0.71 &\textbf{0.71} &0.73 &\textbf{0.73} &0.58 &\textbf{0.71} &\textbf{0.91} &\textbf{0.91} \\\midrule
\textbf{Average} &0.60 &\textbf{0.72} &0.55 &\textbf{0.79} &0.51 &\textbf{0.67} &0.74 &\textbf{0.78} &0.83 &\textbf{0.86} &0.84 &\textbf{0.88} &0.63 &\textbf{0.78} &0.73 &\textbf{0.83} &0.72 &\textbf{0.89} &0.91 &\textbf{0.92} \\
\bottomrule
\end{tabular}
\end{adjustbox}
\end{table*}

\begin{table*}[tp]\centering
\caption{Improvement in PA}\label{tab: pa}
\begin{adjustbox}{width=1\textwidth}
\scriptsize
\begin{tabular}{lrrrrrrrrrrrrrrrrrrrrr}\toprule
  &\multicolumn{2}{c}{\textbf{LFA}}
  &\multicolumn{2}{c}{\textbf{LogCluster}}
  &\multicolumn{2}{c}{\textbf{SHISO}}
  &\multicolumn{2}{c}{\textbf{LogMine}}
  &\multicolumn{2}{c}{\textbf{AEL}}
  &\multicolumn{2}{c}{\textbf{Drain}}
  &\multicolumn{2}{c}{\textbf{UniParser}}
  &\multicolumn{2}{c}{\textbf{LogPPT}}
  &\multicolumn{2}{c}{\textbf{LLMParser}}
  &\multicolumn{2}{c}{\textbf{LILAC}}\\
    \cmidrule{2-21}
      &\textbf{\xmark} &\textbf{\cmark} &\textbf{\xmark} &\textbf{\cmark} &\textbf{\xmark} &\textbf{\cmark}
      &\textbf{\xmark} &\textbf{\cmark} &\textbf{\xmark} &\textbf{\cmark} &\textbf{\xmark} &\textbf{\cmark}
      &\textbf{\xmark} &\textbf{\cmark} &\textbf{\xmark} &\textbf{\cmark} &\textbf{\xmark} &\textbf{\cmark}
      &\textbf{\xmark} &\textbf{\cmark} \\\toprule
      \textbf{Proxifier} &0.00 &\textbf{0.42} &0.00 &\textbf{0.96} &0.50 &\textbf{1.00} &0.00 &\textbf{0.98} &0.68 &\textbf{0.90} &0.69 &\textbf{1.00} &0.63 &\textbf{0.88} &\textbf{1.00} &0.79 &\textbf{0.96} &0.79 &\textbf{1.00} &\textbf{1.00} \\
\textbf{Linux} &0.03 &\textbf{0.63} &0.02 &\textbf{0.64} &0.02 &\textbf{0.58} &0.04 &\textbf{0.61} &0.10 &\textbf{0.61} &0.11 &\textbf{0.63} &0.14 &\textbf{0.66} &0.52 &\textbf{0.64} &0.43 &\textbf{0.59} &0.68 &\textbf{0.69} \\
\textbf{Apache} &0.64 &\textbf{0.80} &0.03 &\textbf{0.99} &0.03 &\textbf{0.99} &0.26 &\textbf{0.99} &0.73 &\textbf{0.99} &0.73 &\textbf{0.99} &0.28 &\textbf{0.99} &0.97 &\textbf{0.99} &\textbf{1.00} &0.99 &0.98 &\textbf{0.99} \\
\textbf{Zookeeper} &0.35 &\textbf{0.81} &0.46 &\textbf{0.82} &0.66 &\textbf{0.80} &0.46 &\textbf{0.82} &\textbf{0.84} &0.83 &\textbf{0.84} &0.83 &\textbf{0.82} &0.82 &\textbf{0.85} &0.82 &\textbf{0.91} &0.82 &0.80 &\textbf{0.82} \\
\textbf{Hadoop} &0.43 &\textbf{0.80} &0.05 &\textbf{0.78} &0.02 &\textbf{0.67} &0.53 &\textbf{0.83} &0.54 &\textbf{0.83} &0.54 &\textbf{0.83} &0.61 &\textbf{0.79} &0.73 &\textbf{0.79} &\textbf{0.83} &0.79 &\textbf{0.87} &0.78 \\
\textbf{HealthApp} &0.31 &\textbf{0.91} &0.02 &\textbf{0.93} &0.01 &\textbf{0.37} &0.31 &\textbf{0.92} &0.31 &\textbf{0.95} &0.31 &\textbf{0.96} &0.46 &\textbf{0.97} &\textbf{1.00} &0.97 &\textbf{0.99} &0.97 &0.57 &\textbf{0.96} \\
\textbf{OpenStack} &0.01 &\textbf{0.65} &0.01 &\textbf{0.84} &0.02 &\textbf{0.87} &0.01 &\textbf{0.97} &0.03 &\textbf{0.73} &0.03 &\textbf{0.81} &0.14 &\textbf{0.98} &0.87 &\textbf{0.98} &0.95 &\textbf{0.98} &0.95 &\textbf{0.98} \\
\textbf{HPC} &0.69 &\textbf{0.86} &0.64 &\textbf{0.89} &0.00 &\textbf{0.13} &\multicolumn{1}{c}{\textemdash} &\multicolumn{1}{c}{\textemdash} &0.74 &\textbf{0.96} &0.72 &\textbf{0.97} &0.75 &\textbf{0.87} &\textbf{1.00} &0.92 &0.41 &\textbf{0.90} &0.99 &\textbf{1.00} \\
\textbf{Mac} &0.23 &\textbf{0.49} &0.12 &\textbf{0.56} &0.11 &\textbf{0.54} &0.28 &\textbf{0.59} &0.24 &\textbf{0.56} &0.36 &\textbf{0.60} &0.31 &\textbf{0.56} &0.48 &\textbf{0.56} &\textbf{0.59} &0.58 &\textbf{0.63} &0.60 \\
\textbf{OpenSSH} &0.05 &\textbf{0.52} &0.03 &\textbf{0.66} &0.14 &\textbf{0.72} &\multicolumn{1}{c}{\textemdash} &\multicolumn{1}{c}{\textemdash} &0.36 &\textbf{0.43} &0.59 &\textbf{0.65} &0.53 &\textbf{0.68} &0.71 &\textbf{0.71} &\textbf{0.88} &0.72 &\textbf{1.00} &0.72 \\
\textbf{HDFS} &0.19 &\textbf{0.76} &0.00 &\textbf{0.92} &0.03 &\textbf{0.94} &\multicolumn{1}{c}{\textemdash} &\multicolumn{1}{c}{\textemdash} &0.46 &\textbf{0.94} &0.46 &\textbf{0.94} &0.81 &\textbf{0.94} &0.90 &\textbf{0.94} &0.86 &\textbf{0.94} &\textbf{1.00} &\textbf{1.00} \\
\textbf{Spark} &0.32 &\textbf{0.79} &0.01 &\textbf{0.82} &0.00 &\textbf{0.53} &\multicolumn{1}{c}{\textemdash} &\multicolumn{1}{c}{\textemdash} &0.33 &\textbf{0.81} &0.33 &\textbf{0.81} &0.36 &\textbf{0.93} &\textbf{0.96} &0.94 &\textbf{0.96} &0.94 &0.76 &\textbf{0.90} \\
\textbf{Thunderbird} &0.06 &\textbf{0.45} &0.08 &\textbf{0.45} &0.06 &\textbf{0.38} &\multicolumn{1}{c}{\textemdash} &\multicolumn{1}{c}{\textemdash} &0.25 &\textbf{0.52} &0.28 &\textbf{0.53} &0.25 &\textbf{0.50} &0.49 &\textbf{0.51} &\textbf{0.72} &0.54 &0.55 &\textbf{0.60} \\
\textbf{BGL} &0.02 &\textbf{0.16} &0.35 &\textbf{0.60} &0.25 &\textbf{0.37} &\multicolumn{1}{c}{\textemdash} &\multicolumn{1}{c}{\textemdash} &0.43 &\textbf{0.61} &0.46 &\textbf{0.62} &0.47 &\textbf{0.67} &\textbf{0.69} &\textbf{0.69} &0.13 &\textbf{0.69} &\textbf{0.98} &\textbf{0.98} \\
\midrule
\textbf{Average} &0.24 &\textbf{0.65} &0.13 &\textbf{0.78} &0.13 &\textbf{0.64} &0.24 &\textbf{0.75} &0.43 &\textbf{0.76} &0.46 &\textbf{0.80} &0.47 &\textbf{0.80} &0.80 &\textbf{0.82} &0.76 &\textbf{0.80} &0.84 &\textbf{0.86} \\
\bottomrule
\end{tabular}
\end{adjustbox}
\end{table*}

\begin{table*}[tp]\centering
\caption{Improvement in FGA}\label{tab: fga}
\begin{adjustbox}{width=1\textwidth}
\scriptsize
\begin{tabular}{lrrrrrrrrrrrrrrrrrrrrr}\toprule
  &\multicolumn{2}{c}{\textbf{LFA}}
  &\multicolumn{2}{c}{\textbf{LogCluster}}
  &\multicolumn{2}{c}{\textbf{SHISO}}
  &\multicolumn{2}{c}{\textbf{LogMine}}
  &\multicolumn{2}{c}{\textbf{AEL}}
  &\multicolumn{2}{c}{\textbf{Drain}}
  &\multicolumn{2}{c}{\textbf{UniParser}}
  &\multicolumn{2}{c}{\textbf{LogPPT}}
  &\multicolumn{2}{c}{\textbf{LLMParser}}
  &\multicolumn{2}{c}{\textbf{LILAC}}\\
    \cmidrule{2-21}
      &\textbf{\xmark} &\textbf{\cmark} &\textbf{\xmark} &\textbf{\cmark} &\textbf{\xmark} &\textbf{\cmark}
      &\textbf{\xmark} &\textbf{\cmark} &\textbf{\xmark} &\textbf{\cmark} &\textbf{\xmark} &\textbf{\cmark}
      &\textbf{\xmark} &\textbf{\cmark} &\textbf{\xmark} &\textbf{\cmark} &\textbf{\xmark} &\textbf{\cmark}
      &\textbf{\xmark} &\textbf{\cmark} \\\toprule
      \textbf{Proxifier} &\textbf{0.40} &\textbf{0.40} &0.00 &\textbf{0.21} &0.36 &\textbf{0.69} &0.01 &\textbf{0.47} &0.52 &\textbf{0.55} &0.21 &\textbf{0.54} &0.05 &\textbf{0.43} &\textbf{0.87} &\textbf{0.87} &0.15 &\textbf{0.46} &\textbf{1.00} &\textbf{1.00} \\
\textbf{Linux} &0.74 &\textbf{0.82} &0.29 &\textbf{0.73} &0.49 &\textbf{0.61} &0.75 &\textbf{0.85} &0.75 &\textbf{0.86} &0.78 &\textbf{0.86} &0.13 &\textbf{0.76} &0.72 &\textbf{0.77} &0.69 &\textbf{0.81} &0.90 &\textbf{0.90} \\
\textbf{Apache} &\textbf{0.87} &\textbf{0.87} &0.00 &\textbf{0.84} &0.78 &\textbf{0.83} &\textbf{1.00} &\textbf{1.00} &\textbf{1.00} &\textbf{1.00} &\textbf{1.00} &\textbf{1.00} &0.26 &\textbf{0.77} &0.68 &\textbf{0.84} &0.73 &\textbf{1.00} &\textbf{1.00} &\textbf{1.00} \\
\textbf{Zookeeper} &0.63 &\textbf{0.71} &0.09 &\textbf{0.84} &0.44 &\textbf{0.62} &0.02 &\textbf{0.85} &0.79 &\textbf{0.90} &\textbf{0.90} &\textbf{0.90} &0.71 &\textbf{0.88} &0.93 &\textbf{0.98} &0.41 &\textbf{0.98} &0.93 &\textbf{0.96} \\
\textbf{Hadoop} &0.65 &\textbf{0.82} &0.00 &\textbf{0.77} &0.55 &\textbf{0.70} &0.12 &\textbf{0.91} &0.12 &\textbf{0.92} &0.79 &\textbf{0.95} &0.63 &\textbf{0.91} &0.60 &\textbf{0.88} &0.41 &\textbf{0.89} &\textbf{0.93} &\textbf{0.93} \\
\textbf{HealthApp} &0.01 &\textbf{0.81} &0.01 &\textbf{0.46} &0.40 &\textbf{0.55} &0.01 &\textbf{0.73} &0.01 &\textbf{0.86} &0.01 &\textbf{0.97} &0.59 &\textbf{0.96} &0.93 &\textbf{0.96} &0.61 &\textbf{0.94} &0.97 &\textbf{0.99} \\
\textbf{OpenStack} &0.60 &\textbf{0.62} &0.00 &\textbf{0.84} &0.53 &\textbf{0.84} &0.00 &\textbf{0.96} &\textbf{0.68} &\textbf{0.68} &0.01 &\textbf{0.71} &0.83 &\textbf{1.00} &\textbf{1.00} &\textbf{1.00} &0.15 &\textbf{1.00} &\textbf{1.00} &\textbf{1.00} \\
\textbf{HPC} &0.35 &\textbf{0.44} &0.05 &\textbf{0.10} &0.12 &\textbf{0.15} &\multicolumn{1}{c}{\textemdash} &\multicolumn{1}{c}{\textemdash} &0.20 &\textbf{0.27} &0.31 &\textbf{0.41} &0.66 &\textbf{0.71} &0.69 &\textbf{0.86} &0.01 &\textbf{0.02} &\textbf{0.96} &\textbf{0.96} \\
\textbf{Mac} &0.65 &\textbf{0.70} &0.07 &\textbf{0.66} &0.44 &\textbf{0.67} &0.45 &\textbf{0.84} &0.79 &\textbf{0.81} &0.23 &\textbf{0.68} &0.70 &\textbf{0.79} &0.48 &\textbf{0.70} &0.20 &\textbf{0.60} &0.87 &\textbf{0.89} \\
\textbf{OpenSSH} &0.51 &\textbf{0.67} &0.01 &\textbf{0.33} &0.61 &\textbf{0.74} &\multicolumn{1}{c}{\textemdash} &\multicolumn{1}{c}{\textemdash} &0.69 &\textbf{0.75} &\textbf{0.87} &\textbf{0.87} &0.02 &\textbf{0.03} &0.14 &\textbf{0.18} &0.37 &\textbf{0.60} &\textbf{0.88} &\textbf{0.88} \\
\textbf{HDFS} &0.86 &\textbf{0.91} &0.00 &\textbf{0.88} &0.95 &\textbf{1.00} &\multicolumn{1}{c}{\textemdash} &\multicolumn{1}{c}{\textemdash} &0.86 &\textbf{0.95} &0.95 &\textbf{1.00} &\textbf{1.00} &\textbf{1.00} &0.74 &\textbf{0.95} &0.57 &\textbf{0.90} &\textbf{1.00} &\textbf{1.00} \\
\textbf{Spark} &0.64 &\textbf{0.79} &0.00 &\textbf{0.23} &0.42 &\textbf{0.54} &\multicolumn{1}{c}{\textemdash} &\multicolumn{1}{c}{\textemdash} &0.02 &\textbf{0.85} &0.89 &\textbf{0.91} &0.25 &\textbf{0.62} &0.48 &\textbf{0.83} &0.18 &\textbf{0.72} &0.86 &\textbf{0.89} \\
\textbf{Thunderbird} &0.71 &\textbf{0.77} &0.01 &\textbf{0.47} &0.56 &\textbf{0.68} &\multicolumn{1}{c}{\textemdash} &\multicolumn{1}{c}{\textemdash} &0.59 &\textbf{0.83} &0.85 &\textbf{0.89} &0.78 &\textbf{0.82} &0.73 &\textbf{0.84} &0.32 &\textbf{0.55} &0.64 &\textbf{0.90} \\
\textbf{BGL} &0.56 &\textbf{0.60} &0.06 &\textbf{0.86} &0.43 &\textbf{0.53} &\multicolumn{1}{c}{\textemdash} &\multicolumn{1}{c}{\textemdash} &0.74 &\textbf{0.78} &0.79 &\textbf{0.82} &0.87 &\textbf{0.92} &0.90 &\textbf{0.95} &0.01 &\textbf{0.02} &\textbf{0.94} &\textbf{0.94} \\
\midrule
\textbf{Average} &0.58 &\textbf{0.71} &0.04 &\textbf{0.59} &0.51 &\textbf{0.65} &0.30 &\textbf{0.78} &0.55 &\textbf{0.79} &0.61 &\textbf{0.82} &0.53 &\textbf{0.76} &0.71 &\textbf{0.83} &0.34 &\textbf{0.68} &0.92 &\textbf{0.95} \\
\bottomrule
\end{tabular}
\end{adjustbox}
\end{table*}

\begin{table*}[tp]\centering
\caption{Improvement in FTA}\label{tab: fta}
\begin{adjustbox}{width=1\textwidth}
\scriptsize
\begin{tabular}{lrrrrrrrrrrrrrrrrrrrrr}\toprule
  &\multicolumn{2}{c}{\textbf{LFA}}
  &\multicolumn{2}{c}{\textbf{LogCluster}}
  &\multicolumn{2}{c}{\textbf{SHISO}}
  &\multicolumn{2}{c}{\textbf{LogMine}}
  &\multicolumn{2}{c}{\textbf{AEL}}
  &\multicolumn{2}{c}{\textbf{Drain}}
  &\multicolumn{2}{c}{\textbf{UniParser}}
  &\multicolumn{2}{c}{\textbf{LogPPT}}
  &\multicolumn{2}{c}{\textbf{LLMParser}}
  &\multicolumn{2}{c}{\textbf{LILAC}}\\
    \cmidrule{2-21}
      &\textbf{\xmark} &\textbf{\cmark} &\textbf{\xmark} &\textbf{\cmark} &\textbf{\xmark} &\textbf{\cmark}
      &\textbf{\xmark} &\textbf{\cmark} &\textbf{\xmark} &\textbf{\cmark} &\textbf{\xmark} &\textbf{\cmark}
      &\textbf{\xmark} &\textbf{\cmark} &\textbf{\xmark} &\textbf{\cmark} &\textbf{\xmark} &\textbf{\cmark}
      &\textbf{\xmark} &\textbf{\cmark} \\\toprule
      \textbf{Proxifier} &0.00 &\textbf{0.40} &0.00 &\textbf{0.21} &0.24 &\textbf{0.69} &0.00 &\textbf{0.47} &0.44 &\textbf{0.55} &0.15 &\textbf{0.54} &0.05 &\textbf{0.43} &\textbf{0.87} &0.70 &0.13 &\textbf{0.31} &\textbf{1.00} &\textbf{1.00} \\
\textbf{Linux} &0.07 &\textbf{0.71} &0.06 &\textbf{0.61} &0.13 &\textbf{0.49} &0.18 &\textbf{0.70} &0.20 &\textbf{0.71} &0.26 &\textbf{0.73} &0.02 &\textbf{0.64} &0.43 &\textbf{0.65} &0.25 &\textbf{0.67} &0.62 &\textbf{0.74} \\
\textbf{Apache} &0.26 &\textbf{0.66} &0.00 &\textbf{0.69} &0.25 &\textbf{0.67} &0.41 &\textbf{0.76} &0.52 &\textbf{0.76} &0.52 &\textbf{0.76} &0.10 &\textbf{0.62} &0.31 &\textbf{0.69} &0.62 &\textbf{0.76} &\textbf{0.83} &0.76 \\
\textbf{Zookeeper} &0.18 &\textbf{0.58} &0.05 &\textbf{0.75} &0.17 &\textbf{0.51} &0.01 &\textbf{0.69} &0.47 &\textbf{0.71} &0.61 &\textbf{0.78} &0.48 &\textbf{0.71} &0.78 &\textbf{0.79} &0.38 &\textbf{0.79} &\textbf{0.82} &0.79 \\
\textbf{Hadoop} &0.10 &\textbf{0.56} &0.00 &\textbf{0.55} &0.06 &\textbf{0.45} &0.06 &\textbf{0.66} &0.06 &\textbf{0.66} &0.38 &\textbf{0.67} &0.47 &\textbf{0.64} &0.46 &\textbf{0.63} &0.28 &\textbf{0.64} &\textbf{0.72} &0.66 \\
\textbf{HealthApp} &0.00 &\textbf{0.71} &0.00 &\textbf{0.41} &0.04 &\textbf{0.46} &0.00 &\textbf{0.63} &0.00 &\textbf{0.75} &0.00 &\textbf{0.86} &0.27 &\textbf{0.85} &0.83 &\textbf{0.85} &0.52 &\textbf{0.83} &0.82 &\textbf{0.88} \\
\textbf{OpenStack} &0.08 &\textbf{0.46} &0.00 &\textbf{0.71} &0.09 &\textbf{0.70} &0.00 &\textbf{0.83} &0.17 &\textbf{0.54} &0.00 &\textbf{0.57} &0.26 &\textbf{0.88} &0.83 &\textbf{0.88} &0.14 &\textbf{0.88} &\textbf{0.90} &0.88 \\
\textbf{HPC} &0.20 &\textbf{0.42} &0.03 &\textbf{0.09} &0.00 &\textbf{0.14} &\multicolumn{1}{c}{\textemdash} &\multicolumn{1}{c}{\textemdash} &0.13 &\textbf{0.26} &0.14 &\textbf{0.40} &0.33 &\textbf{0.66} &0.53 &\textbf{0.81} &0.00 &\textbf{0.02} &0.89 &\textbf{0.92} \\
\textbf{Mac} &0.08 &\textbf{0.44} &0.02 &\textbf{0.45} &0.08 &\textbf{0.44} &0.11 &\textbf{0.56} &0.20 &\textbf{0.53} &0.07 &\textbf{0.45} &0.26 &\textbf{0.52} &0.29 &\textbf{0.47} &0.10 &\textbf{0.40} &0.56 &\textbf{0.59} \\
\textbf{OpenSSH} &0.05 &\textbf{0.58} &0.00 &\textbf{0.31} &0.17 &\textbf{0.69} &\multicolumn{1}{c}{\textemdash} &\multicolumn{1}{c}{\textemdash} &0.31 &\textbf{0.71} &0.44 &\textbf{0.80} &0.01 &\textbf{0.03} &0.13 &\textbf{0.17} &0.31 &\textbf{0.54} &\textbf{0.85} &0.80 \\
\textbf{HDFS} &0.03 &\textbf{0.41} &0.00 &\textbf{0.44} &0.40 &\textbf{0.53} &\multicolumn{1}{c}{\textemdash} &\multicolumn{1}{c}{\textemdash} &0.52 &\textbf{0.54} &\textbf{0.59} &0.53 &\textbf{0.59} &0.53 &0.40 &\textbf{0.52} &0.30 &\textbf{0.55} &\textbf{1.00} &\textbf{1.00} \\
\textbf{Spark} &0.09 &\textbf{0.55} &0.00 &\textbf{0.17} &0.12 &\textbf{0.37} &\multicolumn{1}{c}{\textemdash} &\multicolumn{1}{c}{\textemdash} &0.01 &\textbf{0.61} &0.46 &\textbf{0.65} &0.14 &\textbf{0.41} &0.35 &\textbf{0.55} &0.15 &\textbf{0.51} &\textbf{0.75} &0.63 \\
\textbf{Thunderbird} &0.06 &\textbf{0.49} &0.00 &\textbf{0.30} &0.11 &\textbf{0.40} &\multicolumn{1}{c}{\textemdash} &\multicolumn{1}{c}{\textemdash} &0.18 &\textbf{0.52} &0.28 &\textbf{0.56} &0.31 &\textbf{0.51} &0.44 &\textbf{0.53} &0.22 &\textbf{0.35} &0.44 &\textbf{0.58} \\
\textbf{BGL} &0.03 &\textbf{0.50} &0.01 &\textbf{0.72} &0.07 &\textbf{0.40} &\multicolumn{1}{c}{\textemdash} &\multicolumn{1}{c}{\textemdash} &0.14 &\textbf{0.68} &0.19 &\textbf{0.68} &0.20 &\textbf{0.75} &\textbf{0.79} &0.78 &0.00 &\textbf{0.02} &\textbf{0.90} &\textbf{0.90} \\ \midrule
\textbf{Average} &0.09 &\textbf{0.53} &0.01 &\textbf{0.46} &0.14 &\textbf{0.50} &0.10 &\textbf{0.60} &0.24 &\textbf{0.61} &0.29 &\textbf{0.64} &0.25 &\textbf{0.58} &0.53 &\textbf{0.64} &0.24 &\textbf{0.52} &\textbf{0.79} &\textbf{0.79} \\
\bottomrule
\end{tabular}
\end{adjustbox}
\end{table*}

In this RQ, we apply SynLog+ to all of the benchmark log parsers and evaluate
the efficacy of SynLog+ in improving their accuracy. First, we integrate SynLog+ with each
baseline parser by treating the parser as a grouper that clusters raw log
messages. Second, SynLog+ is applied to each resulting log group to refine the
extracted templates. 

As shown in Table~\ref{tab: pa}, SynLog+ boosts the parsing
accuracy across all baseline syntax-based log parsers. On average, it yields a
236\% improvement in message-level log parsing accuracy of syntax-based log
parsers.  For semantic-based log parsers, the average boost is
lower at 20.6\%.  This is because SynLog+ aims to lift the parsing accuracy
close to the grouping accuracy by operating on the log groups.  Since the
parsing accuracy is already close to the grouping accuracy for the
semantic-based log parsers, SynLog+ does not provide as high a boost as it
provides the syntax-based log parsers for which the parsing accuracy is much
lower than the grouping accuracy. Table~\ref{tab: ga}, \ref{tab: fga}, and \ref{tab: fta} presents the performance
boost in GA, FGA, and FTA, respectively.  SynLog+ achieves 17\% and 181.5\%
increase in GA and FGA, respectively.  For FTA, the syntax-based log parsers
attain an average boost of 831.5\% and the semantic-based parsers attain that of
67\%.


\rqboxc{
  \textbf{O6.} SynLog+ significantly improves the message-level and template-level
  parsing accuracy of all baseline log parsers.  The improvement is most
  noticeable in syntax-based log parsers where the gap between grouping accuracy
  and parsing accuracy were significant.
}

\begin{figure*}[t]
  \centering
  \includegraphics[width=1\textwidth]{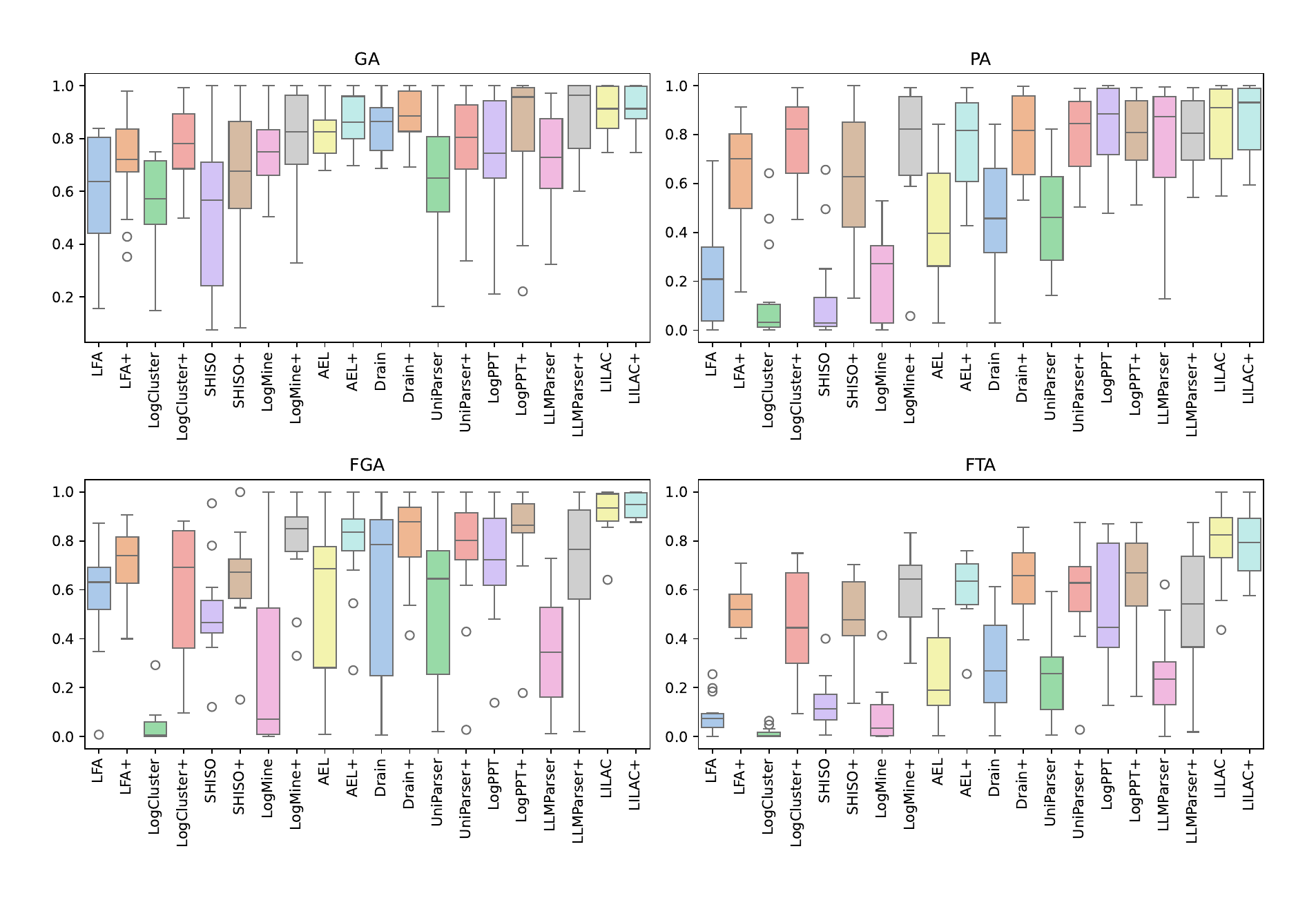}
  \caption{Robustness comparison of SynLog+ with benchmark syntax-based and
  semantic-based log parsers}
  \label{fig: robust}
\end{figure*}

Along with the average accuracy, it is important to compare the robustness of
the log parsers.  If a log parser shows high variance across different log
dataset, that is indicative of low robustness against different logging formats.
Figure~\ref{fig: robust} shows that
not only does SynLog+ improves accuracy across for all baseliens across all four metrics, it also reduces the
variance in the accuracy distribution, demonstrating the robustness of SynLog+
across different log types.

\rqboxc{
  \textbf{O7.} SynLog+ improves the robustness of all baseline log
  parsers across diverse log types.
}

\subsection{RQ7: Efficiency Evaluation}
Besides accuracy, efficiency is another critical metric for log parsers to
consider in order to handle large-scale log data.  To evaluate the efficiency of
SynLog+, we compare its execution time for all 10 baseline log parsers except
LogMine since it is unable to finish parsing these datasets within a reasonable
time limit.  Following RQ2, we evaluate the runtime efficiency using the HPC and
the OpenSSH datasets from the Loghub-2 benchmark~\cite{Loghub2}.


\begin{figure}[t]
  \centering
  \begin{subfigure}{0.49\textwidth}
    \includegraphics[width=\linewidth]{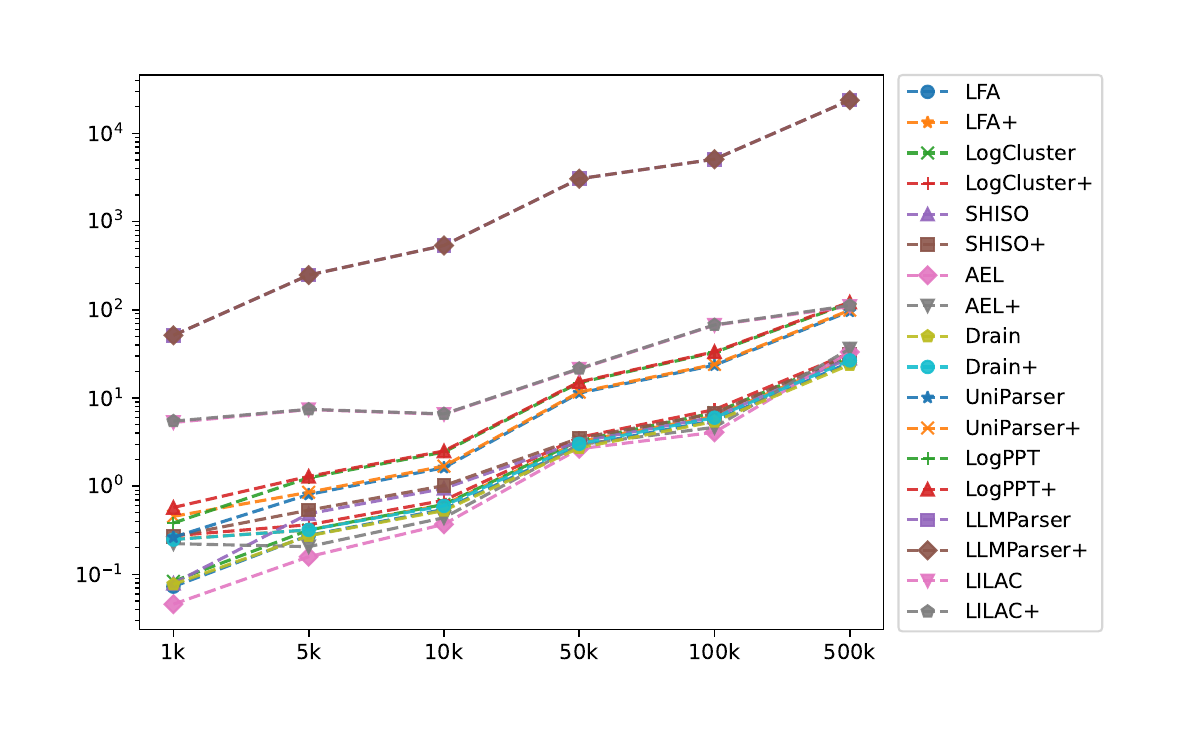}
    \caption{HPC dataset}
  \end{subfigure}
  \begin{subfigure}{0.49\textwidth}
    \includegraphics[width=\linewidth]{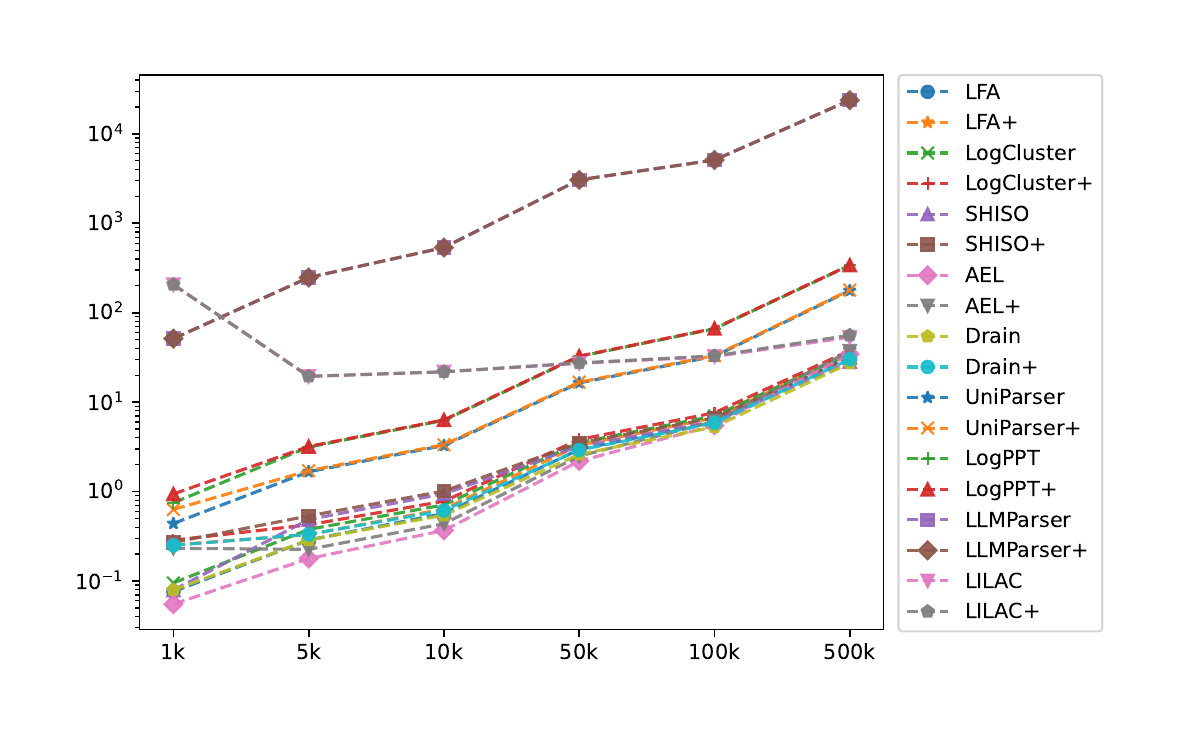}
    \caption{OpenSSH dataset}
  \end{subfigure}
  \caption{Runtime efficiency of SynLog+ when coupled with three of
  the best-performing syntax-based log parsers compared to benchmark syntax-
  and semantic-based log parsers}
  \label{fig: eff}
\end{figure}

Figure~\ref{fig: eff} shows the runtime of the log parsers on HPC and OpenSSH
datasets on a logarithmic scale.  The figure shows that syntax-based log parsers
require significantly smaller execution time compared to the inference time of
the semantic-based log parsers.  Additionally, semantic-based log parsers
require training costs, which are not considered in this efficiency evaluation.
Moreover, when SynLog+ is coupled with the syntax-based log parsers, it incurs
virtually no extra runtime cost, demonstrating its applicability in real-world
large-scale log parsing.

\rqboxc{
  \textbf{O8.}
  SynLog+ introduces virtually no additional runtime cost to syntax-based log
  parsers.  In contrast, semantic-based log parsers require exponentially larger
  execution time, rendering them less practical for large-scale log parsing.
}

\subsection{RQ8: Generalizability of the Regex Patterns}

\begin{table}[t]\centering
  \caption{Frequency analysis of variable types in Loghub-2k}
  \label{tab: regex1}
  \scriptsize
\begin{tabular}{lrrrrrrrrr}\toprule
  \multirow{1.75}{*}{Dataset}
  &\multirow{1.75}{*}{Total} &\multicolumn{1}{c}{MAC} &\multicolumn{1}{c}{Email} &\multicolumn{1}{c}{IP} &\multicolumn{1}{c}{Time} &\multicolumn{1}{c}{Memory} &\multicolumn{1}{c}{Date} &\multicolumn{1}{c}{File} \\
  &\multicolumn{1}{c}{} &\multicolumn{1}{c}{Address} &\multicolumn{1}{c}{Address} &\multicolumn{1}{c}{Address} &\multicolumn{1}{c}{Duration} &\multicolumn{1}{c}{Size} &\multicolumn{1}{c}{\& Time} &\multicolumn{1}{c}{Paths} \\
  \midrule
  Proxifier &5,146 &0 &0 &3,022 &0 &999 &0 &0 \\
Linux &3,901 &0 &0 &1,686 &0 &10 &910 &9 \\
Apache &633 &0 &0 &32 &0 &0 &0 &601 \\
Zookeeper &673 &0 &0 &614 &40 &0 &0 &5 \\
Hadoop &1,928 &0 &0 &73 &2 &0 &0 &4 \\
HealthApp &271 &0 &0 &4 &0 &0 &0 &0 \\
OpenStack &3,270 &0 &0 &847 &0 &60 &0 &378 \\
HPC &214 &0 &0 &93 &0 &0 &0 &0 \\
Mac &1,287 &20 &0 &92 &0 &0 &0 &141 \\
OpenSSH &2,939 &0 &0 &1,826 &0 &0 &0 &0 \\
Spark &922 &0 &0 &0 &0 &298 &0 &1 \\
Thunderbird &1,902 &40 &0 &654 &0 &17 &1 &55 \\
BGL &220 &10 &0 &36 &0 &1 &0 &119 \\
HDFS &2,806 &0 &0 &750 &0 &0 &0 &715 \\
Android &1,503 &0 &0 &88 &3 &0 &0 &0 \\
Windows &637 &0 &0 &8 &0 &0 &0 &6 \\\midrule
Total &28,252 &70 &0 &9,825 &45 &1,385 &911 &2,034 \\
\bottomrule
\end{tabular}
\end{table}

\begin{figure}[t]
    \centering\begin{tikzpicture}[scale=0.3]
    \scriptsize
    \pie[
        /tikz/every pin/.style={align=center},
        text=pin, number in legend,
        explode=0.0,
        radius=5,
            color={black!0, black!10, black!20, black!30, black!40},
            ]
            {
                19.2/\textbf{MemSize} \\ (19.2\%),
                18.6/\textbf{Object Id} \\ (18.6\%),
                 2.0/\textbf{Username} \\ (2.0\%),
                59.4/\textbf{Others} \\ (59.4\%),
                 0.8/\textbf{DataSrc} \\ (0.8\%)
            }
     \end{tikzpicture}
	\caption{Distribution of variables unmached by regex patterns}
	\label{fig: unmatched}
\end{figure}
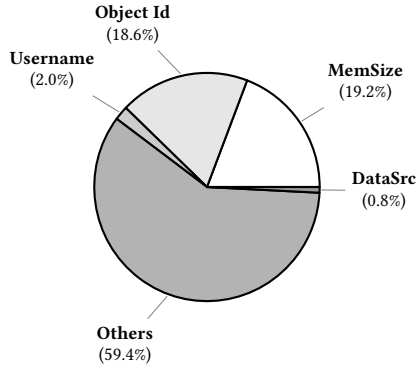

To answer the question of generalizability of the regex patterns, we explore the
Loghub-2k benchmark \cite{Loghub}, which consists of 16 datasets each with 2k
logs.

Li et al. categorized log variables into 10 variable types. SynLog+ uses the
regular expression patterns presented in Table~\ref{tab: reganon} to match these
variable types. The frequency of each variable type presented in Table~\ref{tab:
regex1} show that 50\% of the variables can be identified with the proposed
regex patterns. In Apache and Zookeeper, more than 90\% of the variables are
file paths. Moreover, IP addresses and file paths together cover 50\% of the
variables in total.

Figure~\ref{fig: unmatched} presents the distribution of the 50\% variables that
remain unmatched by SynLog+ regex patterns.  38\% of them are memory sizes
(`1038 bytes (1.03 KB)') and object identifiers (`blk\_7236'), whose patterns
are domain- or dataset-specific. Usernames (`cyrus', `aadmin1') and data sources
(`cn602') consists 2.7\% of the unmatched variables.  The rest of them are of no
distinguishable patterns.

\rqboxc{
  \textbf{O9.}
  The regex patterns used by SynLog+ are able to capture 50\% of the variables
  in 16 Loghub-2k datasets, demonstrating the generalizability of the patterns
  across a variety of log types.
}

\subsection{RQ9: Sensitivity Analysis}

\begin{table}[t]\centering
  \caption{Sensitivity Analysis}\label{tab: sensitivity}
  \scriptsize
    \begin{tabular}{lrrrrrrrrrrrrr}\toprule
      \multirow{2.5}{*}{\textbf{Dataset}}
&\multicolumn{4}{c}{\textbf{k = 2}} &\multicolumn{4}{c}{\textbf{k = 3}} &\multicolumn{4}{c}{\textbf{k = 6}} \\\cmidrule{2-13}
&\textbf{GA} &\textbf{PA} &\textbf{FGA} &\textbf{FTA} &\textbf{GA} &\textbf{PA} &\textbf{FGA} &\textbf{FTA} &\textbf{GA} &\textbf{PA} &\textbf{FGA} &\textbf{FTA} \\\midrule
\textbf{Proxifier}    &0.69 &1.00 &0.54 &0.54 &0.69 &1.00 &0.54 &0.54 &0.69 &1.00 &0.54 &0.54 \\
\textbf{Linux}        &0.71 &0.63 &0.86 &0.73 &0.71 &0.63 &0.86 &0.73 &0.71 &0.63 &0.86 &0.73 \\
\textbf{Apache}       &1.00 &0.99 &1.00 &0.76 &1.00 &0.99 &1.00 &0.76 &1.00 &0.99 &1.00 &0.76 \\
\textbf{Zookeeper}    &0.99 &0.83 &0.90 &0.78 &0.99 &0.83 &0.90 &0.78 &0.99 &0.83 &0.90 &0.78 \\
\textbf{Hadoop}       &0.94 &0.83 &0.95 &0.67 &0.94 &0.83 &0.95 &0.67 &0.94 &0.83 &0.95 &0.67 \\
\textbf{HealthApp}    &0.99 &0.96 &0.97 &0.86 &0.99 &0.96 &0.97 &0.86 &0.99 &0.96 &0.97 &0.86 \\
\textbf{OpenStack}    &0.82 &0.81 &0.71 &0.57 &0.82 &0.81 &0.71 &0.57 &0.82 &0.81 &0.71 &0.57 \\
\textbf{HPC}          &0.84 &0.97 &0.41 &0.40 &0.84 &0.97 &0.41 &0.40 &0.84 &0.97 &0.41 &0.40 \\
\textbf{Mac}          &0.90 &0.60 &0.68 &0.45 &0.90 &0.60 &0.68 &0.45 &0.90 &0.60 &0.68 &0.45 \\
\textbf{OpenSSH}      &0.71 &0.65 &0.87 &0.80 &0.71 &0.65 &0.87 &0.80 &0.71 &0.65 &0.87 &0.80 \\
\textbf{HDFS}         &1.00 &0.94 &1.00 &0.53 &1.00 &0.94 &1.00 &0.53 &1.00 &0.94 &1.00 &0.53 \\
\textbf{Spark}        &0.87 &0.81 &0.91 &0.65 &0.87 &0.81 &0.91 &0.65 &0.87 &0.81 &0.91 &0.65 \\
\textbf{Thunderbird}  &0.87 &0.53 &0.89 &0.56 &0.87 &0.53 &0.89 &0.56 &0.87 &0.53 &0.89 &0.56 \\
\textbf{BGL}          &0.91 &0.62 &0.82 &0.68 &0.91 &0.62 &0.82 &0.68 &0.91 &0.62 &0.82 &0.68 \\
\midrule
\textbf{Average}      &0.86 &0.82 &0.79 &0.65 &0.86 &0.82 &0.79 &0.65 &0.86 &0.82 &0.79 &0.65 \\
\bottomrule
\end{tabular}
\end{table}

To increase efficiency, SynLog+ samples a set of representative logs to identify
the constants and variables instead of analyzing all the logs in a log group. We
evaluate the performance of SynLog+ combined with Drain with $k=2$, $k=3$, and
$k=6$, where k is the number of representative logs selected from each group. As
presented in Table~\ref{tab: sensitivity}, the experiments resulted in the same
accuracy scores across all four metrics.  This demonstrates that there is no
gain in accuracy by increasing the number of representative logs.

\rqboxc{
  \textbf{O10.}
  Increasing the number of representative logs does not result in an increased
  accuracy.  Thus, it is preferable to use a small number of representative logs
  for constant and variable identification for greater efficiency.
}

\begin{table*}[t]
  \centering
  \caption{Recommendations for log parsers based on study findings (Ob = Observation)}
  \scriptsize
  \begin{tabular}{lp{0.60\textwidth}l}\toprule
    {\textbf{ID}} & \textbf{Recommendation} & \textbf{Rationale w/ Ob} \\
    \midrule
    R1 &Syntax-based methods should be preferred for efficient and accurate log
    grouping &O1, 2\\
    R2 &Semantic-based methods should be preferred for accurate
    template identification &O1\\
    R3 &Semantic-based log parsers needs to be generalizable &O3\\
    R4 &Two-phase log parsing should be preferred over single-phase log parsing &O4\\
    R5 &Log grouping should be done before template identification &O3, 5, 6, 7\\
    R6 &Leverage domain-agnostic heuristics (e.g., regex patterns) &O6, 7, 8, 9 \\
    \bottomrule
  \end{tabular}%
  \label{tab: recom}
\end{table*}

\section{Recommendations}

Based on the study findings of log parsing techniques and architectures
presented in this paper, and the proposed two-phase log parsing pipeline
SynLog+, we present the following recommendations in Table~\ref{tab: recom}.

\textbf{R1. Favour syntax-based methods for efficient and accurate log
grouping.} RQ1 and RQ2 have shown that syntax-based log parsing techniques
provide better grouping accuracy (O1) and better efficiency (O2). Semantic-based
log parsers, on the other hand, obtain better parsing accuracy but incur 10 to
1,000 times more runtime costs (O2). Given the high throughput requirements of
large-scale log analysis systems, we recommend that log parsers favor
syntax-based techniques for matching incoming logs to existing templates instead
of invoking the semantic model for every new log message.

\textbf{R2. Favour semantic-based methods for accurate template identification.}
RQ1 and RQ2 have also shown that semantic-based log parsers are better at
accurately identifying the log templates (O1).  Although this accuracy comes at
the cost of a decreased efficiency (O2), log parsers should favour semantic-based
methods to accurately identify the constant and variable tokens of the log templates.

\textbf{R3. Study on generalizable semantic-based log parsers.}  Semantic-based
log parsers suffer a hit in accuracy when evaluating their performance only on
unseen log data (O3).  Semantic-based models require a subset of log data to
train on.  Overfitting on these training data can cause the performance to
appear inflated.  Further research should be conducted to increase the generalizability
of the semantic-based log parsers.

\textbf{R4. Adopt two-phase architecture.}  The experimental results indicate
that decoupling the grouping and template identification steps can improve
accuracy, obtaining 27\%, 49\%, and 81\% improvement in GA, FGA, and FTA,
respectively (O4).  LILAC, the baseline log parser with the highest accuracy, is
a two-phase log parser. Moreover, the two-phase architecture of SynLog+ improves
the template accuracy of syntax-based log parsers by 236\% on average without
using a semantic model (O6).  Based on these observations, we recommend that log
parsers employ two-phase log parsing architecture.

\textbf{R5. Log grouping before template identification.} LILAC achieves the
highest accuracy among all the baseline log parsers.  The cause behind LILAC's
impressive accuracy is its two-phase architecture, specifically its template
cache acting as the log grouping phase followed by an LLM acting as the template
identification phase. Moreover, SynLog+ acts as a two-phase pipeline where a
grouper is used to group the log messages and then a template identification
module is used to identify the constant and variable tokens.  Based on the
effectiveness of LILAC and SynLog+ (O1, O6), we recommend that log parsers
Perform log grouping before template identification.

\textbf{R6. Leverage domain-agnostic heuristics.} While recent log parsers have
moved away from regex-based pre-processing due to concerns over domain
specificity, Li et al. have shown that log variables often fall into a limited
set of categories. Notably, three of these categories can be identified with
regex pattern matching, such as IP addresses, time/duration, memory size or
clock cycles, etc. The success of SynLog+ (O6, O7) shows that heuristics such as
regular expressions can provide substantial gains at minimal cost (O8).
Experimental results show that 50\% of the variables in Loghub-2k can be
identified by general-purpose regex patterns used by SynLog+ (O9). Log parsers
should first consider rule-driven pre-processing before resorting to complex
models.

\section{Conclusion}

In this paper, we conducted an empirical study which analyzes the performances
of syntax-based and semantic-based log parsers in addition to the impact of
grouping and template identification phases in two-phase log parsing
architectures.  The results of the study demonstrate that syntax-based log
parsers achieve better grouping accuracy but are outperformed by semantic-based
log parsers in parsing accuracy.  Additionally, the study shows that compared
to single-phase architecture, two-phase architecture obtains better accuracy
and efficiency.

Additionally, we introduced SynLog+, a lightweight regex-based template
extractor which extracts the log template from the log groups clustered by a
syntax-based log parser.  SynLog+ improves the parsing accuracy of
syntax-based log parsers while retaining their grouping accuracy.  SynLog+
coupled with the syntax-based log parser Drain achieves highest average
template-level accuracies, outperforming the semantic-based log parsers.

\bibliographystyle{ACM-Reference-Format}
\bibliography{refs}

\end{document}